\documentclass{aa}
\usepackage{lscape}
\usepackage{graphicx}
\usepackage{txfonts}
\usepackage{natbib}
\bibpunct{(}{)}{;}{a}{}{,} 
\usepackage[colorlinks, citecolor=blue]{hyperref}
\usepackage{changepage}
\usepackage{fixltx2e}
\def\Planck{\textit{Planck}}

\begin{document}

\title{Characterization of a subsample of the \Planck\ SZ source cluster catalogues using optical SDSS DR12 data}

\titlerunning{Validation of \Planck\ SZ source cluster catalogues}

\author{A. Streblyanska \inst{1,2} \and
R. Barrena \inst{1,2}  \and
J.A.~Rubi\~{n}o-Mart\'{\i}n \inst{1,2} \and
R.F.J.~van der Burg\inst{3} \and
N. Aghanim \inst{4} \and
A.~Aguado-Barahona \inst{1,2} \and 
A.~Ferragamo\inst{1,2} \and
H. Lietzen \inst{5} 
}
 
\institute{Instituto de Astrof\'{\i}sica de Canarias, C/V\'{\i}a L\'{a}ctea s/n, La Laguna, Tenerife, Spain\\ 
     \email{alina@iac.es} \and
 Universidad de La Laguna (ULL), Dept. Astrof\'{\i}sica, E-38200 La Laguna, Tenerife, Spain \and
IRFU, CEA, Universit{\'e} Paris-Saclay, F-91191 Gif-sur-Yvette, France \and
Institut d'Astrophysique Spatiale, Universit\`e Paris-Sud, CNRS, UMR8617, 91405 Orsay, France \and
Tartu Observatory, Faculty of Science and Technology, University of Tartu, Observatooriumi 1, 61602, T\"{o}ravere, Estonia
}

\date{Received ; accepted }

\authorrunning{Streblyanska et al. }

\abstract{}
{The \Planck\ catalogues of Sunyaev-Zeldovich (SZ) sources, PSZ1 and PSZ2, are the largest catalogues of galaxy
clusters selected through their SZ signature in the full sky. In 2013, we started a long-term observational program 
at Canary Island observatories with the aim of validating $\sim$500 unconfirmed SZ sources. In this work we present 
results of the initial pre-screening of possible cluster counterparts using photometric and spectroscopic data 
of the Sloan Digital Sky Survey DR12. Our main aim is to identify previously unconfirmed PSZ2 cluster candidates and to
contribute in determination of the actual purity and completeness of \Planck\ SZ source sample.}
{Using the latest version of the PSZ2 catalogue, we select all sources overlapping with the SDSS DR12 footprint and without 
redshift information. We validate these cluster fields following optical criteria (mainly distance with respect to the 
\Planck\ pointing, magnitude of the brightest cluster galaxy and cluster richness) and combining them with the 
 profiles of the \Planck\ Compton $y$-maps. Together, this procedure allows for a more robust identification of optical 
 counterparts compared to simply cross-matching with existing SDSS cluster catalogues that have been constructed from earlier SDSS Data Releases.}
{The sample contains new redshifts for 37 \Planck\ galaxy clusters that were not included in the original release of PSZ2  \Planck\ catalogue .
We detect three cases as possible multiple counterparts. We show that a combination of all available information
(optical images and profile of SZ signal) can provide correct associations between the observed \Planck\ SZ source and the optically identified cluster. 
We also show that \Planck\ SZ detection is very sensitive even to high-z ($z>$ 0.5)
clusters. In addition, we also present updated spectroscopic information for 34 \Planck\ PSZ1 sources (33 previously 
photometrically confirmed and 1 new identification).}
 {}


\maketitle

\begin{figure*}[ht!]
\centering
\includegraphics[width=15cm]{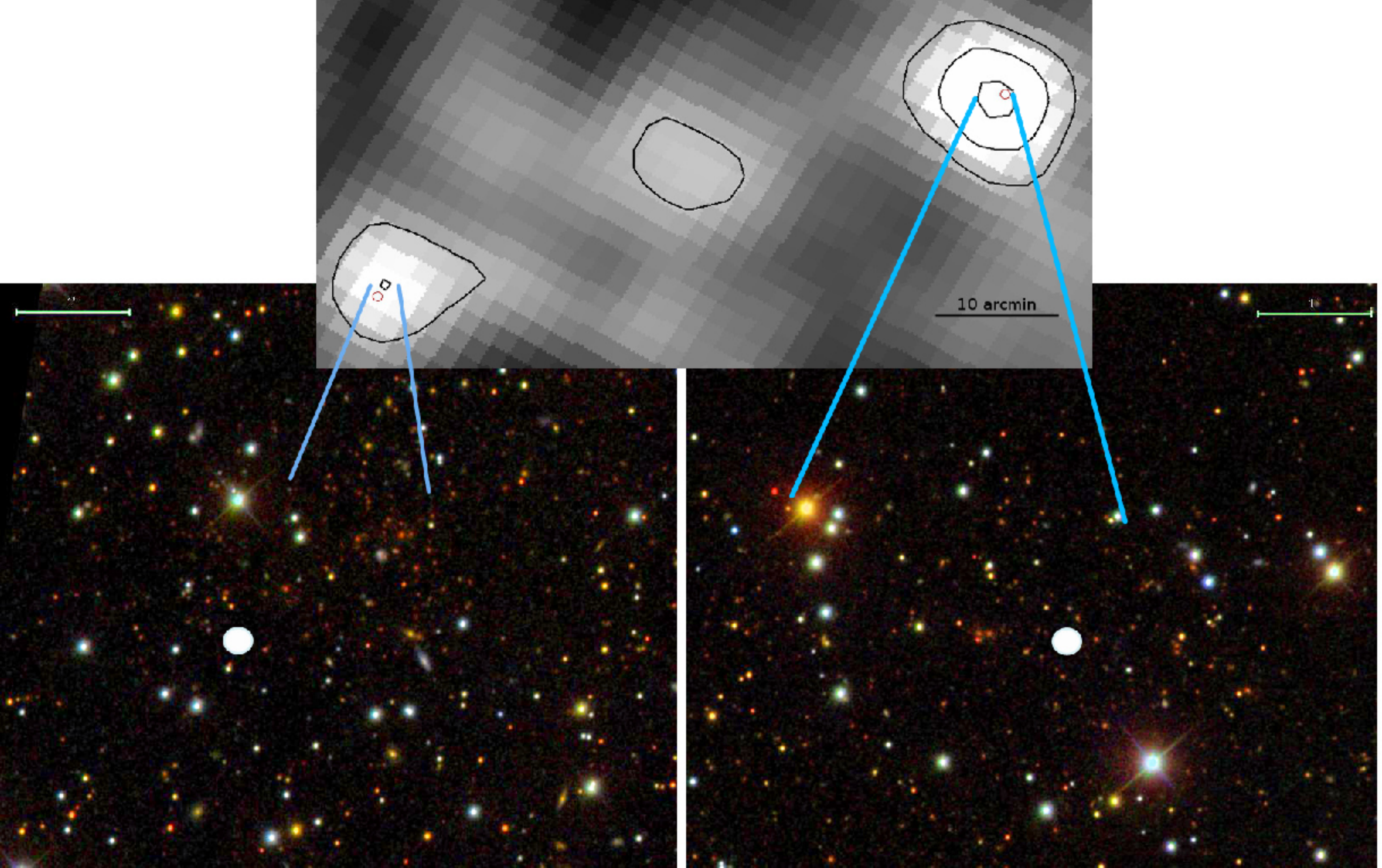}
\caption{The MILCA Compton $y$-map of the area around two confirmed SZ sources, PSZ2 G066.34$+$26.14 
(left, confirmed in this paper, SZ SNR=5.6) and PSZ2 G066.41$+$27.03 (right, validated in \citet{planck2014-XXVI}, SZ SNR=8.8), superimposed to 
a zoomed RGB SDSS images. The black contours correspond to the 3, 6, 9 $\times$10$^{-6}$ levels of the Compton $y$-map 
in this area. The single black contour in the middle of the $y$-map corresponds to a noise fluctuation. The red circles 
mark the nominal PSZ2 source locations. In the RGB SDSS images this nominal position is marked with white filled circles. 
Both PSZ2 sources show the presence of clusters at similar high-z $z_{\rm phot}=$0.62 and 0.53, respectively. Blue lines
delimit the area corresponding to the peak of SZ signal for PSZ2 G066.34$+$26.14 which enclosed almost all cluster 
members. Scale bars in RGB images are 1$\arcmin.$ 
}

\label{fig:psz2_277}
\end{figure*}

\section{Introduction}

Massive galaxy clusters are an excellent tool to test cosmological models and to constrain cosmological parameters, e.g., dark matter or 
dark energy densities \citep[e.g.,][]{vikhlinin09,planck2013-p15}. In the last decade,
the Sunyaev-Zeldovich (SZ) effect \citep{sz1972} is being used as a powerful technique to detect galaxy clusters. This
effect produces a spectral distortion of the cosmic microwave background (CMB) generated by the inverse Compton 
interaction between the CMB photons and the hot intracluster gas of electrons. Using this technique, the 
\Planck\footnote{\Planck\ \url{http://www.esa.int/Planck} is a project of the European Space Agency (ESA) with 
instruments provided by two scientific consortia funded by ESA member states and led by Principal Investigators 
from France and Italy, telescope reflectors provided through a collaboration between ESA and a scientific consortium 
led and funded by Denmark, and additional contributions from NASA (USA).} satellite provided for the first time the 
possibility of detecting galaxy clusters via the SZ effect in a full sky survey \citep{planck2013-p05a,planck2014-a36}.

The early SZ (ESZ) catalogue presented 189 clusters detected from the first 10 months of survey data 
\citep{planck2011}. In 2013, the \Planck\ Collaboration released the first official version of the SZ catalogue, the 
PSZ1, which included 1227 sources detected during the first 15.5 months of observations \citep{planck2013-p05a}. 
Finally, in 2015, the second catalogue PSZ2 \citep{planck2014-a36} presented a complete list from the full-mission 
survey of 29 months. It contains 1653 sources (716 of them are unique for PSZ2). Even today, the PSZ2 catalogue is the largest 
SZ-selected sample of galaxy clusters with full-sky coverage. 

The SZ surface brightness is independent of the redshift, but no information on the source's redshift can be obtained from the signal. Yet, the abundance of clusters at high-z are very sensitive
to the cosmology \citep{borgani01}, so the identification of the counterparts of SZ clusters at other wavelengths
are absolutely mandatory to make these catalogues scientifically 
useful. In this context, a few extensive follow-up programmes of the \Planck\ Collaboration have been dedicated to 
confirm SZ sources using on-earth and space facilities \citep{planck4, planck2015-XXXVI, planck2014-XXVI}.

Telescopes at Canary Islands Observatories are actively participating in the validation and characterization efforts of the ESZ and PSZ1 
catalogues through the International Time Programme\footnote{ITP: \url{http://www.iac.es/eno.php?op1=5&op2=13&lang=en}} (ITP12-2 and ITP13-08)
\citep[e.g.,][]{planck2015-XXXVI, barrena18}. 
Since 2015, we started the second 2-year long-term follow-up programme (LP15) at the Observatorio del Roque de los Muchachos to 
confirm new PSZ2 \Planck\ cluster candidates and to measure their redshifts and other physical properties. The current paper is
a first publication from the serie of articles decicated to this follow-up campaign. In this paper, 
we present the results of the initial pre-screening of possible targets for this follow-up using the SDSS DR12
photometric and spectroscopic data available for PSZ2 targets. If a cluster counterpart is confirmed in the SDSS
data, new imaging observations were not required in our LP15 programme and the cluster was directly considered for spectroscopy, with
the aim of obtaining its mean redshift, velocity dispersion and dynamical mass. Whenever SDSS spectra were also 
available for several cluster members, we provide the spectroscopic redshift of the cluster based on those SDSS
data and we considered it for a more detailed dynamical analysis. If the cluster were not confirmed through SDSS images, we included the 
cluster in our list of targets to retrieve deep imaging information in order to check the presence or absence of high-z 
counterparts. 

This paper is organized as follows. Sections~\ref{sec:sample} and \ref{sec:sdss} describe the \Planck\ PSZ2 catalogue 
and SDSS data, respectively. Section \ref{sec:new} describes our methodology used for new cluster identification.
In Section~\ref{sec:disc}, we discuss the results of the PSZ2 characterization here exposed. Section~\ref{sec:psz1} provides 
updated spectroscopic information for PSZ1 sources. Conclusions are presented in Section \ref{sec:conclusions}.

\begin{figure*}[ht!]
\centering
\includegraphics[width=15cm]{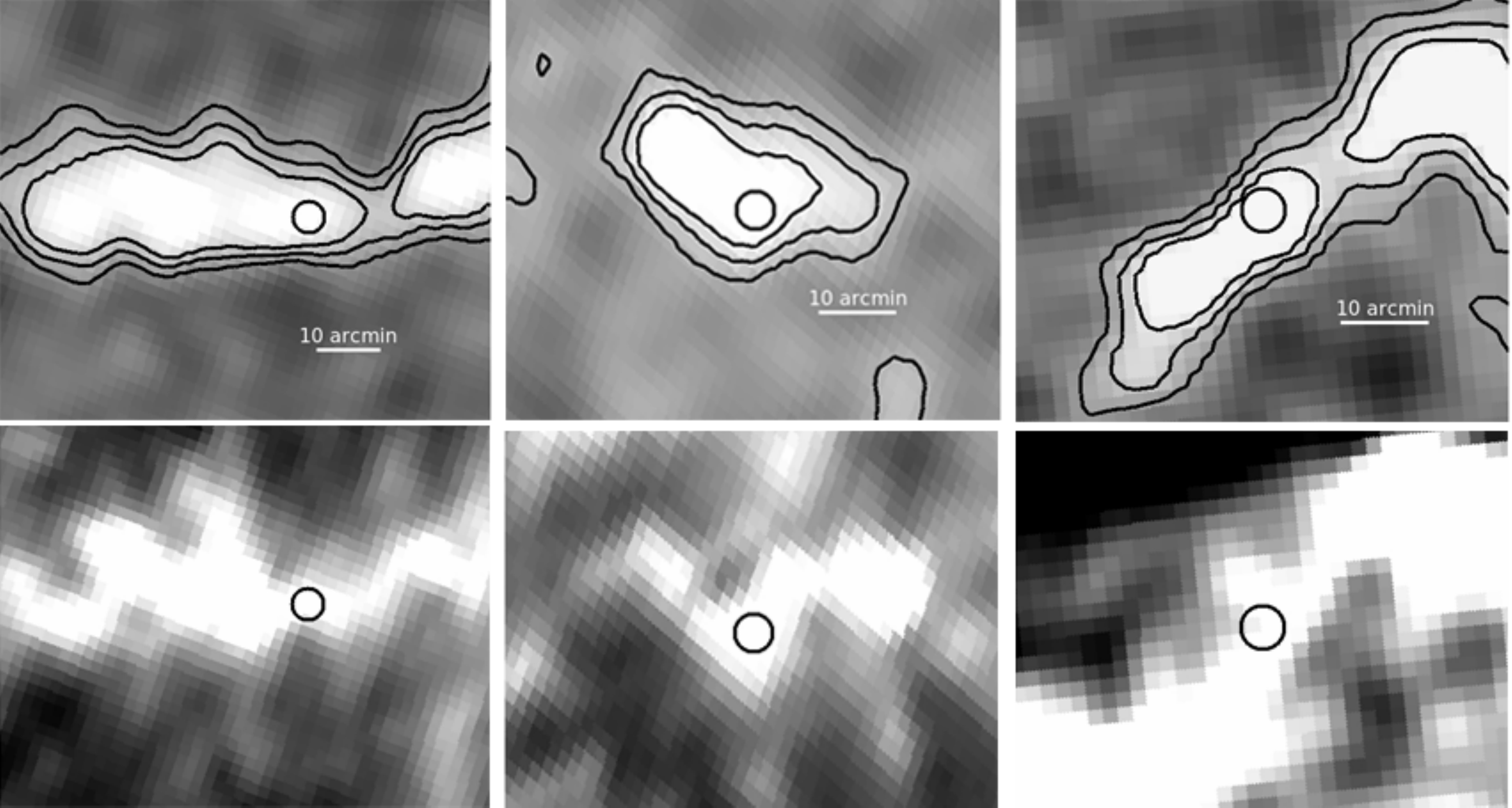}
\caption{Three examples of \Planck\ clusters where the SZ signal seems to be strongly influenced 
by Galactic dust contamination. From left to right three columns correspond to PSZ2 G029.87-17.81, PSZ2 G148.60-48.61 and 
PSZ2 G179.33-22.22, respectively. We present for each source Compton $y$-map (top row) together with the corresponding \Planck\  857 GHz map (bottom row). 
These clusters have SNR$\sim$5 in the PSZ2 catalogue. Black contours correspond to the 3, 6 and 12$\times$10$^{-6}$ 
levels of the Compton $y$-maps in this area. No obvious peak at the nominal position of \Planck\ sources (showed as black
circles) is evident in these three cases. The 857 GHz map for each source suggests that most of the emission arises from dust. 
The SZ flux profile follows an elongated structure that coincided with the direction of a dust Galactic filament.  While there exist the possibility that cluster counterparts correspond to
high-z clusters, in the absence of deeper data we classify them as false SZ detections.}
\label{fig:psz2_non}
\end{figure*}

\section{The \Planck\ PSZ2 cluster sample}
\label{sec:sample}


The SZ detection algorithms applied to make the PSZ2 catalogue are extended and refined 
versions of those used to construct the PSZ1 \citep{planck2013-p05a}. 
As in the case of PSZ1 sources, the PSZ2 cluster candidates are blindly selected using three different detection methods: two 
multifrequency methods based on the matched filter (MMF1 and MMF3) and the PowellSnakes (PwS) procedure, which is based on a fast Bayesian approach to
discrete object detection. The main PSZ2 catalogue contains all objects found by at least one of these three methods 
with a S/N$>$4.5. The PSZ2 catalogue and the techniques used to
construct it are described in detail in \citet{planck2014-a36}. Some additional restrictions were
imposed in order to avoid the spurious detections from the PwS algorithm and to remove detections that were confirmed to be 
spurious by the PSZ1 follow-up. In summary, the PSZ2 catalogue contains 1653 detections (937 sources are common to PSZ1, while 716 are new; 
291 of the PSZ1 catalogue are not in PSZ2). 
1203 of these objects are confirmed clusters and 1094 have redshift estimates.  Confirmation and redshift 
were obtained  from PSZ1,  cross-correlation with other catalogues (e.g. MCXC, ACT, SPT, RedMapper) and from 
the recent specific follow-up campaigns of different international teams  \citep[for detailed description 
see Sec.7 in][]{planck2014-a36}. The main difference in the optical identification of the PSZ1 and PSZ2 sources in the SDSS region is the use of 
 {\tt redMaPPer}  information \citep[][]{rykoff14,rozo15}. 374 sources were cross-identified with a cluster in the v5.10 {\tt redMaPPer} catalog, while further search of counterpart in SDSS data with the {\tt redMaPPer} 
 algorithm  confirmed 17 additional sources (see Section~\ref{sec:sdss}).

\section{Previous SZ follow-up using SDSS data}

\label{sec:sdss}

The SDSS data covers most of the northern extragalactic sky and provides optical images and spectra between 3800-9200 
\AA. Several cluster catalogues have been constructed using different data releases and detection algorithms \citep[][]{hao10, wen12}. The optical validation of SZ sources in the PSZ1 catalogue has been done mainly through the 
cross-correlation with the cluster catalogue by \citet{wen12}, which used the DR8 version of the SDSS data 
\citep{2011ApJS..193...29A}. However, the {\tt redMaPPer} cluster catalogue \citep{rykoff14}, using the DR8 release as 
well, contains many more clusters and has proven to be very efficient at optically identifying PSZ1, and latter PSZ2 
clusters.

The method based on the {\tt redMaPPer} algorithm detects clusters by selecting spatial overdensities of galaxies 
belonging to the red-sequence of the cluster candidate. This procedure is performed in a $10'$ matching radius and imposing some 
constraints relative to the richness and the angular separation between sources. The algorithm provides photometric 
redshift information and richness estimates for all detected clusters. The newly released SDSS-based {\tt redMaPPer} catalogue (v5.10)
produced 374 high-quality matches where the counterparts are consistent with \Planck\ mass and position information.
More than half of these clusters are common for other catalogues, like WHL SDSS sources \citep{wen12} or RXC X-ray sources 
\citep{pi11}. However,  for 95 new PSZ2 Planck sources, the {\tt redMaPPer} catalogue provided the most robust external confirmation 
(redshift and position of the optical cluster).


Although the {\tt redMaPPer} algorithm is proven to be very efficient, as all automatic methods, it has its own limitations. For 
example, low richness systems, in particular nearby clusters (0.05$<$z$<$0.2) are missed due to the high thresholds for richness estimates. 
Additionally, distant clusters ($z>$0.5) could be missed due to problems with photometric errors of the faint galaxies. 
In fact, the PSZ2 catalogue includes 17 confirmed clusters for which the automated {\tt redMaPPer} search fails, but 
individual case-by-case inspection yields real counterparts.

For all these reasons, before starting the observational programme to optically validate unknown PSZ2 sources at Canary 
Islands observatories, an initial pre-screening of the target list in the SDSS database searching for new optical 
counterparts is absolutely necessary. We present in this work the results of this pre-screening with the advantage 
that, nowadays, using the DR12\footnote{\url{http://www.sdss.org/dr12/}} data release we manage a larger sky coverage 
and many more galaxy spectra than in previous releases. Therefore, the new information here presented includes two well defined parts. First, an 
optical validation of unconfirmed PSZ2 sources, and second, an update of the redshift information (going from photometric estimates to spectroscopic 
measurements) for the PSZ1 clusters.

\begin{figure}[ht!]
\centering
\includegraphics[width=8.5cm]{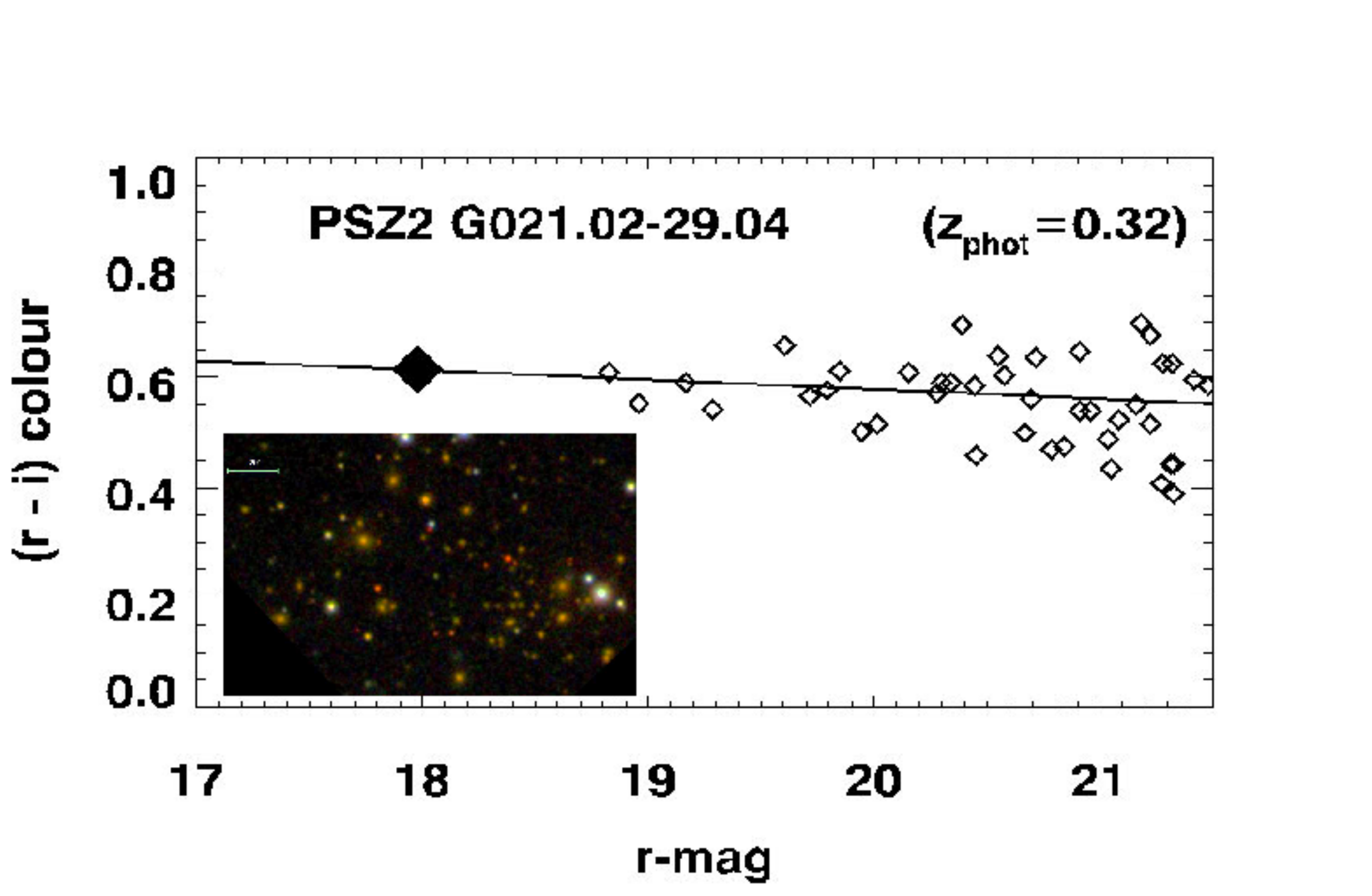}
\caption{Example of the colour-magnitude diagram using data for cluster PSZ2 G021.02-29.04. The red sequence (solid line) is consistent with a photometric 
redshift of 0.32 according to the relation of \citet{planck2015-XXXVI}. 
We consider galaxies with a colour within $\pm0.05$ from the red sequence as likely cluster members.
}
\label{fig:cmd}
\end{figure}

 \begin{figure*}[ht!]
\centering
\includegraphics[width=7.9cm]{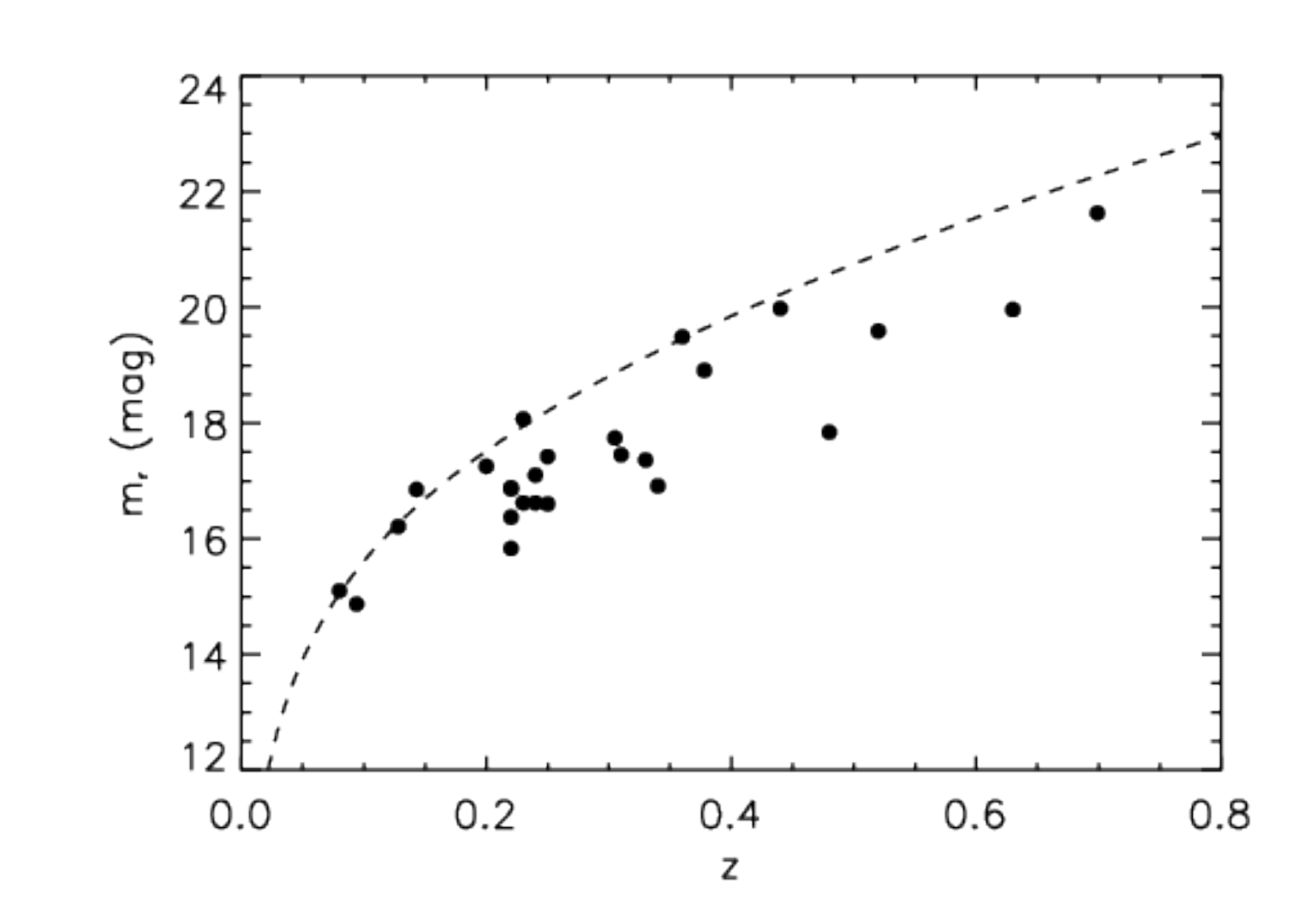}
\includegraphics[width=7.9cm]{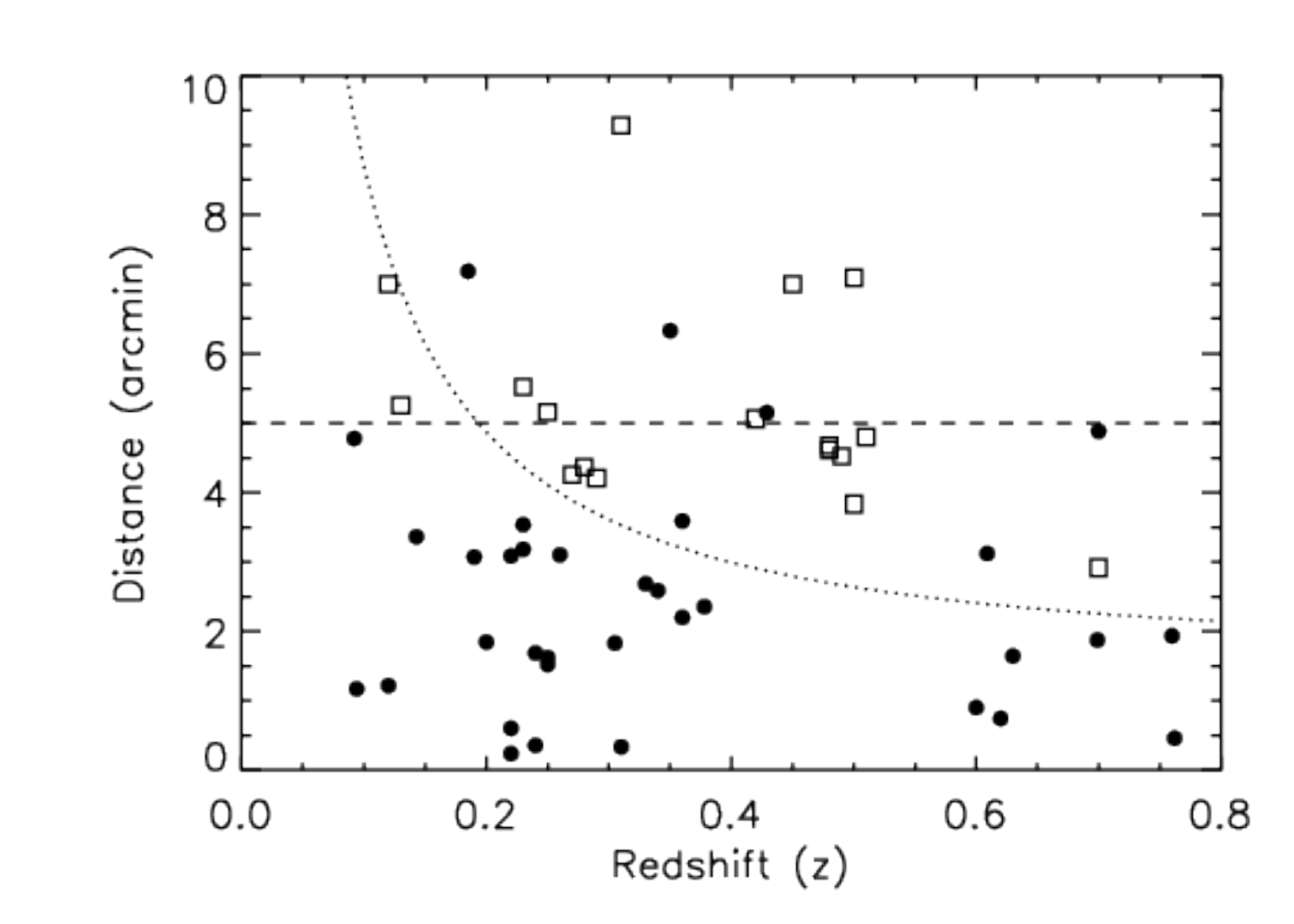}
\caption{Left: Apparent BCG $r$-band magnitude versus redshift of the complete sample of 54 sources studied in this work. The dashed curve 
corresponds to the simple prescription $M_r^\ast -1.5$. Right: Cluster optical centre offsets relative to their \Planck\ SZ position as a function 
of cluster redshift for a sample of 51 sources. Cases with multiple optical counterparts have been excluded from
this analysis. The dashed horizontal line is 5$\arcmin$, which represents the maximum offset 
expected for a \Planck\ SZ detection (i.e. $2\times$FWHM \Planck\ multifrequence combined beam). The dotted line corresponds 
to the physical 1 Mpc radius region at the corresponding redshift. The black dots corresponds to a sample 
of 34 clusters from Table \ref{tab:newpsz2}. These sources fulfill our distance and richness constraints (see 
Sec. \ref{sec:new}). The open squares correspond to 17 clusters which were classified as 
"potential optical counterparts" (Table \ref{tab:newpsz2_add}). }
\label{fig:sdss_z3}
\end{figure*}


\section{Methodology for the validation and characterization of PSZ2 sources}

\label{sec:new}


The cross-correlation between positions of unconfirmed PSZ2 sources and the SDSS survey yields 114 SZ cluster 
candidates covered in the DR12 footprint. In order to validate these clusters, we design a four step criterium detailed 
in the following subsections.

\subsection{Visual inspection on an individual basis}

Using the SDSS DR12 Navigate tool\footnote{\url{skyserver.sdss.org/dr12/en/tools/chart/naviinfo.aspx}} we retrieve a
$10 \arcmin \times 10 \arcmin$ image centred at the nominal location of the \Planck\ SZ source. We visually inspect the image 
looking for possible galaxy overdensities and the presence of a bright cluster galaxy (BCG).  An image with a field
of view of $10 \arcmin \times 10 \arcmin$ is large enough to cover the expected uncertainty in the \Planck\ detection,
which is a FWHM of $2\arcmin-6 \arcmin$ \citep[see Fig. 3 in][]{planck2015-XXXVI}. Clusters and rich groups at 0.1$<z<$0.4 can 
be easily identified in the SDSS RGB images as a concentration of galaxies of the same colour. This first step, the visual 
inspection, is particularly important in the detection of clusters with $z>0.5$. The SDSS photometric survey is 95\% complete 
at a magnitude limit of $r$'=22.2. This means that only BCGs and a few additional galaxy members will be visible in clusters 
at $z>0.5$. In addition, faint galaxies present larger photometric errors and thus redshift uncertainties.
Therefore, distant clusters can be easily missed by automatic algorithms, but the ``inspection by eye'' can easily 
detect these systems.

Some SDSS images show problems with the presence of very bright stars, which make the automatic detection of sources almost impossible.
In some cases, the SDSS database does not supply photometric information, mainly because the field is located in the edge of the surveys. Due to
these facts, we removed 10 sources from the initial list of 114 cluster candidates. These cluster fields are
considered to be observed by our follow-up program (LP15) with the Canary Islands telescopes.

\subsection{Inspection of Compton y-maps}
The full-sky Compton parameter maps ($y$-map) obtained by the \Planck\ Collaboration using the {\tt NILC} and {\tt MILCA} procedures 
\citep{planck2014} are publicly available since 2015 \citep{planck2015}. These maps, constructed from linear combinations of the individual 
\Planck\ frequency charts, preserve the SZ signal and cancel the influence of the CMB. These maps have been widely
exploited in the literature. For example, they are used to obtain an accurate measurement of the SZ power spectrum at intermediate and 
large angular scales \citep{planck2015}, or to study the SZ effect using higher order statistical estimators, such as the skewness and the 
bispectrum. The $y$-map is also used to extract the SZ signal on regions centred at cluster positions and, in 
particular, to perform stacking analyses \citep{planck2015}. These $y$-maps also reveal the diffuse emission at the outskirts of 
clusters and in bridges between merging systems \citep{planck2013c}. Moreover, $y$-maps show SZ signatures 
even for poor galaxy clusters and they help in the confirmation of cluster candidates from PSZ1 catalogue \citep[][]{planck2015-XXXVI, barrena18}.

\begin{figure}[ht!]
\centering
\includegraphics[width=7.7cm]{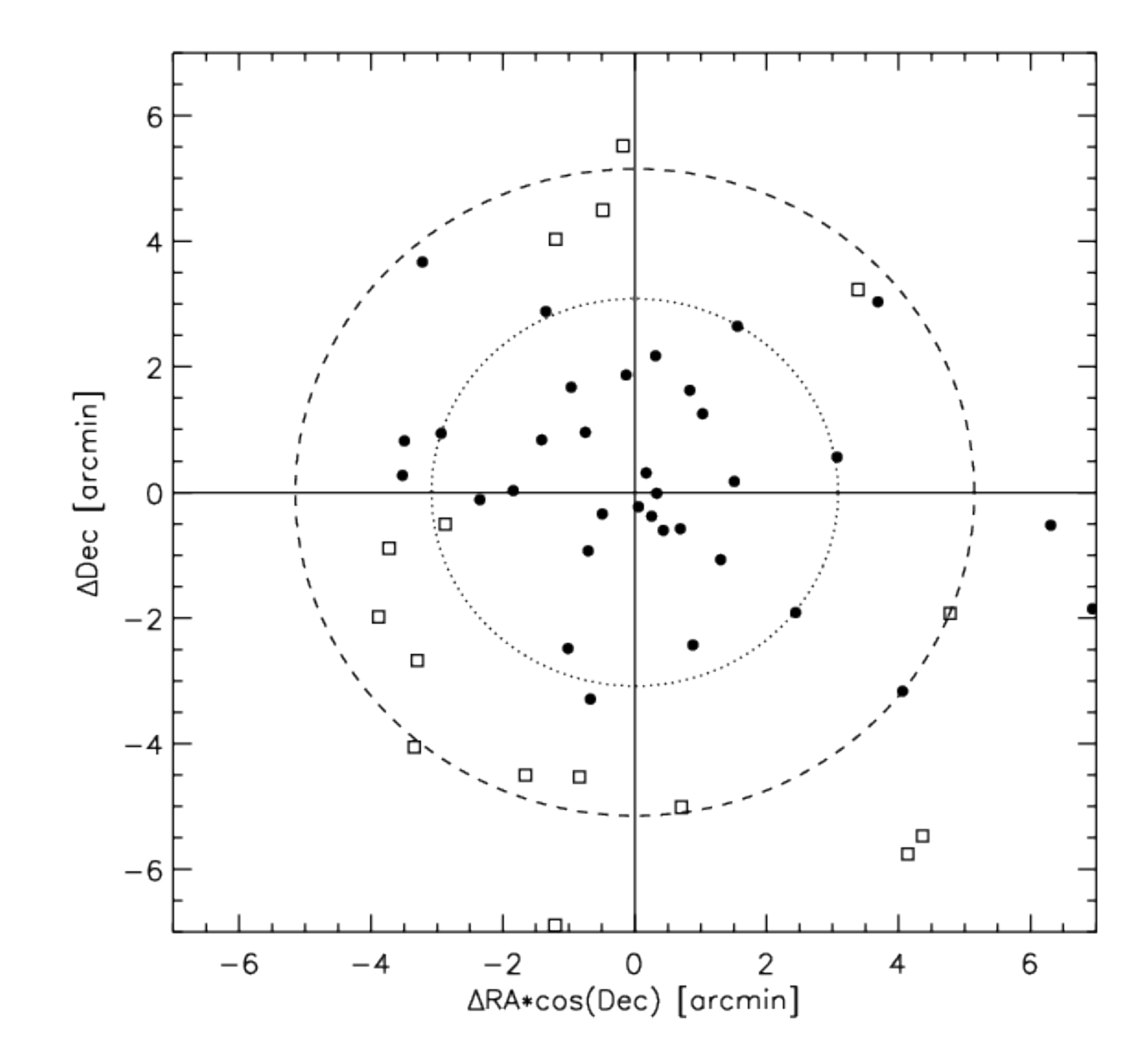}
\caption{Distribution of the optical centre offsets relative to their \Planck\ SZ positions for the sample of 51 
 clusters. 34 confirmed clusters included in Table \ref{tab:newpsz2} (dots), and 17
clusters potentially associated to the corresponding PSZ2 source listed in Table \ref{tab:newpsz2_add} (squares).
Inner dashed line corresponds to 3$\arcmin$.1 radius region, which encloses the 68\% of the
PSZ2 confirmed clusters. External dashed line encloses the $2\times$FWHM beam of the
\Planck\ SZ detection. Cases with multiple optical counterparts have been excluded from this analysis.
}
\label{fig:sdss_z2}
\end{figure}

In Fig.~\ref{fig:psz2_277} we present two typical examples of coincidences between the observed SZ profile and optical images for rich optical 
clusters. The $y$-map shows two clear peaks well above the noise level corresponding to two real optical clusters 
located at angular distance of $\sim 50\arcmin$. One is the already-confirmed cluster PSZ2 G066.41$+$27.03 \citep[WHL J269.219+40.13,][]{planck2014-XXVI}.
After exploration of the SDSS images we could confirm the second source, PSZ2 G066.34$+$26.14.

Although $y$-maps are a valuable source of information for validation works, they must be used with care. It is known 
that the SZ reconstruction might be influenced by the Galactic dust radio emission. In some cases, dust emission may enhance the SZ signal, creating spurious detections. Such situations 
can be partially controlled by the measurements of Galactic extinction in the considered area, and by visual examination 
of the optical (mainly $g-$band) images and $y$-map shapes. In case of strong contamination by Galactic dust, the 
distribution of the signal in the $y$-maps is not compact and shows elongated profiles without a clear peak around the nominal 
position of the \Planck\ SZ source. In many cases, the elongation of the $y$-map contours agrees well with the direction of 
Galactic dust structures visible in the $g-$band image. \citet{barrena18}, \citet{planck2015-XXXVI} and \citet{burg16}  confirmed this fact by 
classifying a large amount of PSZ1 targets as "non-detections", most
of them located at low Galactic latitudes and showing high SZ significance, even S/N $>7$ in some cases (for example 
PSZ1 G071.64$-$42.76). Fig.~\ref{fig:psz2_non} present three examples of $y$-maps for the PSZ2 targets where no optical counterparts
have been identified. The possibility of counterparts associated with high redshift clusters is not excluded, but only deeper 
optical images would confirm or reject this fact. However, the irregular contour profiles observed in $y$-maps and its morphological coincidence with the
structure observed at 857 GHz (thermal dust emission) suggest that the most probable scenario is a false SZ detection.

Taking into account all facts outlined above, we have inspected the $y$-map (obtained through the {\tt MILCA} algorithm) of each PSZ2 
source paying special attention to the profile shape, SZ signal and intensity above the noise level ($>$3$\times$10$^{-6}$
in units of the corresponding $y$-map).
 
\begin{figure*}[ht!]
\centering
\includegraphics[width=8.2cm]{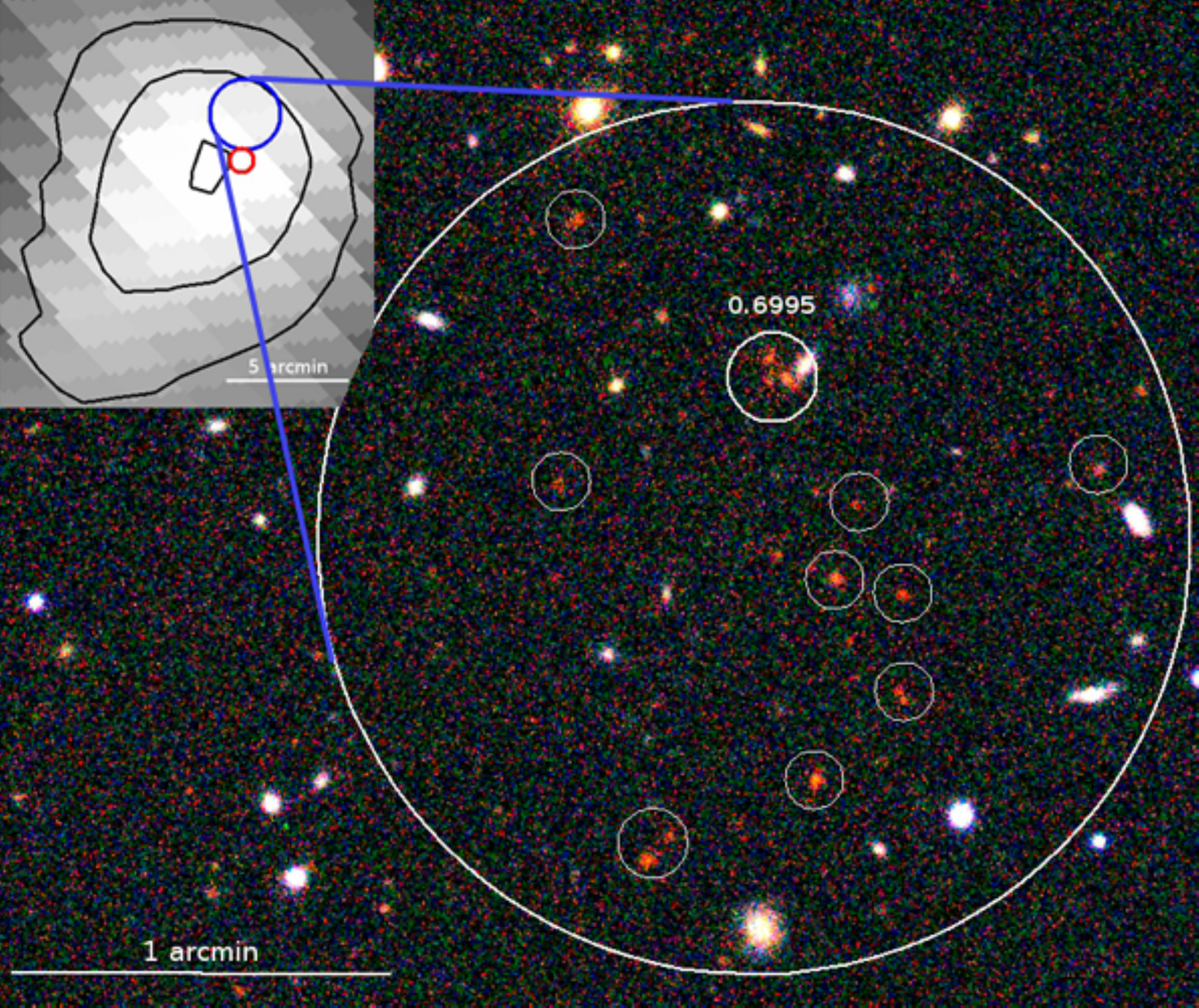}
\includegraphics[width=7.5cm]{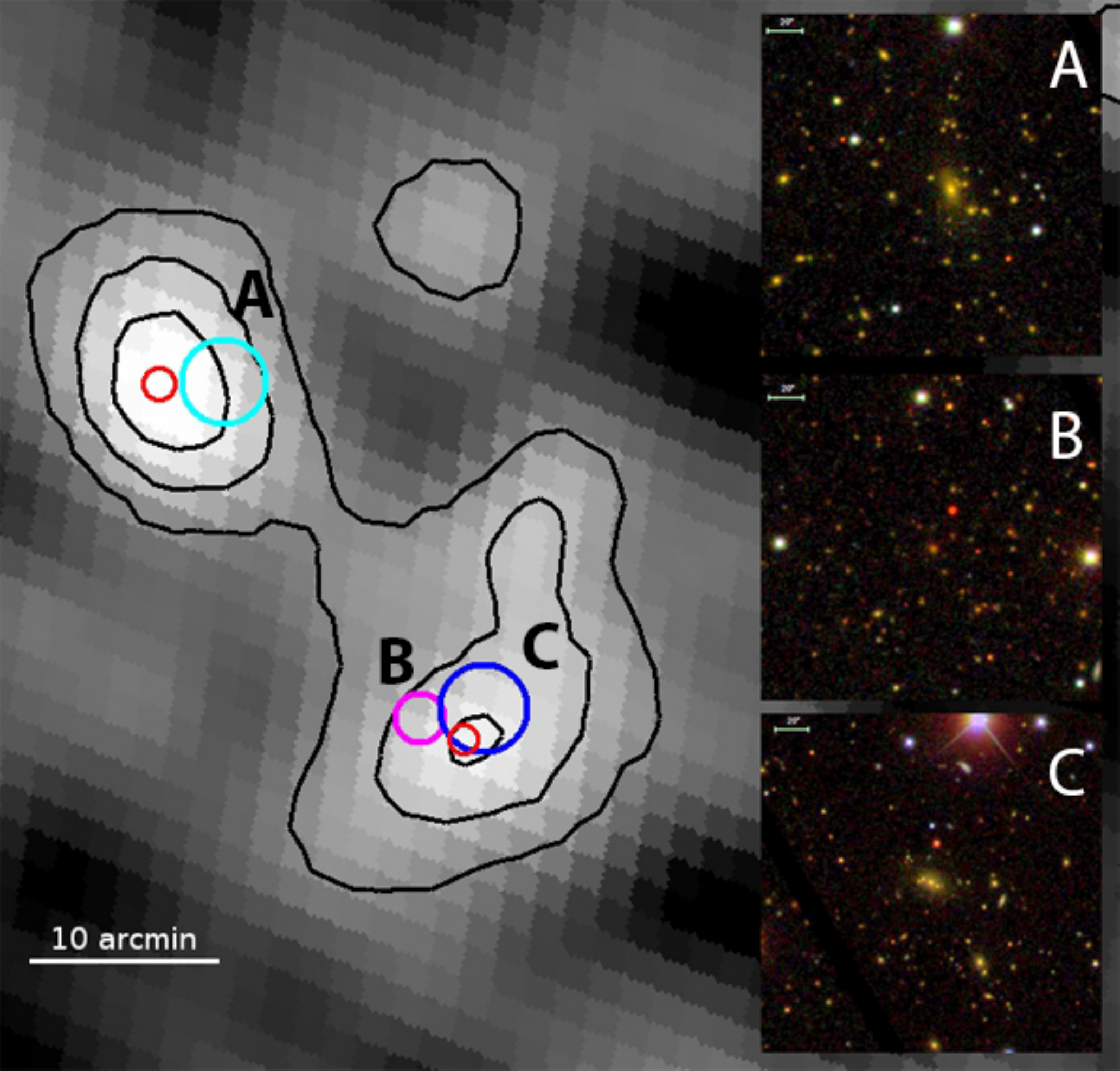}

\caption{Left: Compton $y$-map and RGB SDSS image of PSZ2 G086.28$+$74.76. The black contours correspond 
to the 3, 5 and 6$\times$10$^{-6}$ levels of the Compton $y$-map in this area. The red circle marks the nominal PSZ2 position.
Big white circle overlapping the RGB image corresponds to 1 Mpc at $z=$0.7. We found that the SZ signal comes 
from a cluster at $z_{\rm spec}$ of 0.699. We find 12 likely galaxy members at $z_{\rm phot}\sim$0.75$^{+0.07}_{0.08}$ (small circles). The BCG is at $z_{\rm spec}$=0.6995.
Right: PSZ2 G270.78+36.83 (bottom) and PSZ2 G270.88+37.23 (top) are two  sources separated by the distance of about 25$\arcmin$. We identify one counterpart for PSZ2 G270.88+37.23 at $z_{\rm phot} \sim$0.23 
(cyan circle) and two counterparts for PSZ2 G270.78+36.83 at $z_{\rm phot}\sim$ 0.52 and $z_{\rm phot} \sim$0.22 (magenta and blue circles), 
respectively. The red small circles mark the nominal PSZ2 source coordinates and the size of the circles corresponds to 1 Mpc at the cluster 
redshift. We insert the zoomed RGB images of the centres of these 
clusters. Scale bar in the RGB images is 20$\arcsec$.}
\label{fig:psz2_381}
\end{figure*}

\subsection{Cluster red sequence}

We retrieve SDSS DR12 photometry for each field and the photometric redshift information of galaxies using SQL 
Search\footnote{\url{skyserver.sdss.org/dr12/en/tools/search/sql.aspx}}. We inspect colour-magnitude diagrams (see Fig.\ref{fig:cmd})
($g^\prime - r^\prime$,$r^\prime$) and ($r^\prime - i^\prime$,$i^\prime$) looking for the cluster-red sequence 
(CRS) \citep[e.g.,][]{2000AJ....120.2148G}. 
We fit the red sequence considering all galaxies with colours in the range $\pm0.05$ respect the colour of the BCG. 
This allows us to derive the photometric redshift of the galaxy overdensities following prescriptions detailed in \citet{planck2015-XXXVI}, Sec. 4.2. 
After identifying the galaxy clumps and deriving their $z_{\rm phot}$, we evaluate the richness of these systems in order to validate the galaxy overdensities as actual SZ counterparts.
We detail this process in the following subsection.

\subsection{Richness}
The SZ clusters of the \Planck\ catalogue are massive structures \citep[see e.g. Fig. 28, 29 in][]{planck2013-p05a}. Assuming
that cluster "richness" can be understood as "optical mass tracer" \citep[e.g.,][]{rozo2009b}, galaxy clusters associated with \Planck\ 
SZ emission should also present rich galaxy cluster populations. So, in order to validate our SZ detection, we can estimate 
the richness of the galaxy population in our optical cluster sample.

\citet{rozo2009a,rozo2009b} propose a robust richness estimators from optical catalogues assuming a matched filter model for 
the density profile of galaxies. This method is part of the {\tt redMaPPer} algorithm and proposed to be applied in 
photometric catalogues which sample the galaxy population of the systems. However, our sample includes clusters at 
intermediate redshift, where only the brightest galaxies are visible, making it very difficult to find confidence models to 
the galaxy density profiles. In addition, we consider SDSS photometry, which is not a very deep photometric survey. 

A simple prescription to estimate the richness of a cluster is to compute the actual number of cluster galaxies in a magnitude 
range within the virialized region. For a given cluster with a photometric redshift $z_{phot}$, we count likely cluster  members 
(assumed as galaxies 
within $\pm 0.05$) of the $z_{phot}$ of the cluster showing $r$- magnitudes in the range $(r_{BCG}, r_{BCG}+2.5)$ within the 0.5 Mpc from 
the cluster centre. For this task we consider as a BCG the most
luminous galaxy inside the identified likely cluster members. 
In order to obtain the background subtracted richness and decontaminate from the galaxy field contribution, we compute 
galaxy counts in a $10 \times 10 $ degrees region (centred at RA=10:28:00, DEC=55:20:00) where SDSS DR12 photometric sample presents an 
homogeneous total coverage. We apply the same procedure as for the cluster region 
(assuming the same $z_{phot}$ and $r$- magnitudes ranges), thereby creating a "field galaxy sample" or "local background" for each 
cluster. This field sample was scaled to the 0.5 Mpc area and subtracted from the cluster counts  to obtain a statistical 
estimate of the number of galaxies for each cluster.

However, one of the main problems to estimate the richness in clusters at medium-high redshift ($z>0.3$) is the magnitude 
limit of the photometric sample. Given that SDSS DR12 sample is 95\% complete\footnote{\url{http://www.sdss.org/dr12/scope/}} 
down to $r$'=22.2, we restrict this study to $z<0.6$. Beyond this redshift the $r_{BCG}+2.5$ magnitude is higher than 
the magnitude limit $r$=22.2. Another source of underestimation of richness is the faintness of the galaxy clusters at
medium-high redshift. Possible errors in $z_{phot}$ estimates of faint galaxies may introduce large uncertainties in the
cluster counts, and therefore making a robust richness estimate very difficult. Nonetheless, we consider the richness an
indicative parameter in order to distinguish between the real and false counterparts at $z<0.6$. 

The mass-redshift distribution of the \Planck\ SZ sources is described in \citet{planck15}, which provides the detectable mass of SZ sources in 
the PSZ1 catalogue. Following this relation, no poor systems could be considered actual counterparts even if they are well aligned with the 
\Planck\ pointing. However, the \Planck\ CMB and $y$-maps contain noise, making it possible that a significant fraction of low-mass haloes 
scatter over the SZ detection threshold. This effect is known as Eddington bias \citep{edd13}. \citet{burg16} analyze this effect and confirm 19 SZ clusters at $z>$0.5 with 
M$_{500} \sim 2 \times 10^{14}$ M$_\odot \ h_{70}^{-1}$. Moreover, nearby ($z<0.2$) galaxy systems may also be detected in the \Planck\ SZ maps 
even if they have masses M$_{500} \sim 10^{14}$ M$_\odot \ h_{70}^{-1}$. 
Clusters with lower masses are not expected to be actual SZ counterparts.

In order to find an appropriate threshold to separate real from false counterparts, we study the richness (as mass tracer) of two poor galaxy 
clusters. These are Abell 725 and Abell 796, at $z=$0.092 and 0.157, which present global velocity dispersions of 
$\sigma_v = 534 \pm 115$~km\,s$^{-1}$ and $586 \pm 160$~km\,s$^{-1}$, respectively \citep{boschin2008}.
The mass estimates of these systems show M$_{500} \sim 3 \times 10^{14}$M$_\odot \ h_{70}^{-1}$, so they are low mass 
clusters. By calculating the richness, as defined above, we find that Abell 725 presents 8 cluster members
within a region of 0.5 Mpc radius in the magnitude range $(r_{BCG}, r_{BCG}+2.5)$. In the same way, Abell 796 presents
a richness of 5 galaxies. Therefore, taking into account these numbers, we can conclude that clusters can be considered 
actual SZ counterparts if they present a richness at least of 5 galaxies, according to our richness definition. 

In order to make our sample selection uniform and taking into account the large photometric errors for faint distant sources, 
we assumed this richness cut for high-z clusters as well. However, we noticed that most of our high-z clusters ($z>$ 0.4) have R higher than 5, 
following the expected trend.


\subsection{Spectroscopic data}

In addition to the photometric information, we also use the SDSS DR12 spectroscopic information available around the 
\Planck\ PSZ2 source positions. We also use the SDSS-III photometric cluster catalogue by \citet{wen12}, the group galaxy 
catalogue constructed by \citet{tempel14} of low-redshift candidates, and the DR12 BOSS-CMASS galaxy list for the 
high redshift clusters. We complement this information with some spectroscopic results obtained with the validation programs \citep{planck2015-XXXVI}. 
Given that PSZ2 counterparts show a few galaxy members with spectroscopic redshifts, the 
spectroscopic information was only used to determine the exact redshift of the cluster. 



\begin{figure}[ht!]
\centering
\includegraphics[width=8.cm]{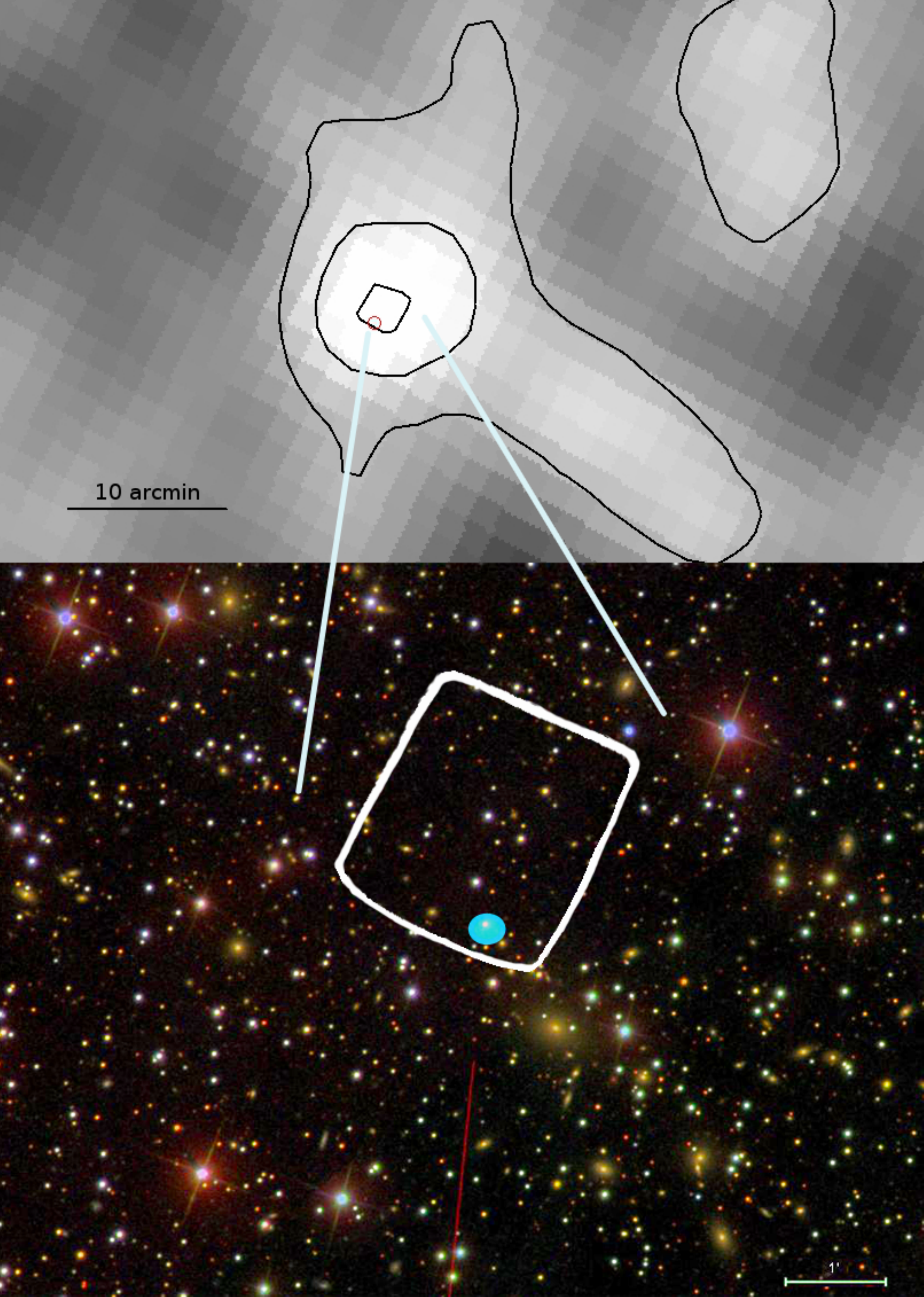}
\caption{Compton y-map (MILCA) and zoomed RGB SDSS image of the region around PSZ2 G069.35$-$15.58. The black contours correspond to the 3, 6, 
9$\times$10$^{-6}$ levels of the Compton y-map in this area where the small red circle marks the nominal \Planck\ position from the PSZ2 catalogue. 
In the RGB image, the white contour corresponds to the 9$\times$10$^{-6}$ level and the PSZ2 coordinates. The BCG of this system is clearly placed at less than $1\arcmin$ to the south west from the nominal \Planck\ pointing. Potential cluster members are consistent with a $z_{\rm phot} \sim 0.09$ and occupy the 
area of $\sim$ 12$\arcmin \times$ 12$\arcmin$.}
\label{fig:psz2_293}
\end{figure}

\section{Results for the PSZ2 catalogue}
\label{sec:disc}

Our validation and characterization procedure is based on the simultaneous use of optical images and Compton $y$-maps. By analyzing 
already-confirmed PSZ1 and PSZ2 sources, we find that, in most cases, there is a strong dependence between the location of the optical 
counterpart and the intensity, shape and size of the SZ signal in the $y$-maps especially for high SNR clusters. This dependence could be extrapolated to 
still unconfirmed sources, and could help to confirm less rich and less concentrated clusters.

Using this combined approach, we were able to identify optical counterparts for 54 new \Planck\ PSZ2 cluster candidates. 

In Fig.~\ref{fig:sdss_z3} (left plot) we present the apparent $r$-band magnitudes of the Bright Central Galaxies (BCGs) versus redshift of 
the complete sample of 54 sources studied in this work. Following \citet{men09}, the dashed curve corresponds to the simple prescription 
$M_r^\ast -1.5$, and assuming passive evolution with 
redshift. The values for $M_r^\ast$ are taken from \citet{bla03}, and the effective wavelengths for each SDSS filter are 
taken from Table 1 in \citet{bla07}. As expected, all the identified BCGs identified show  
magnitudes brighter than $M_r^\ast -1.5$, which represents a necessary condition for a correct cluster identification.

\begin{figure}[ht!]
\centering
\includegraphics[width=8.5cm]{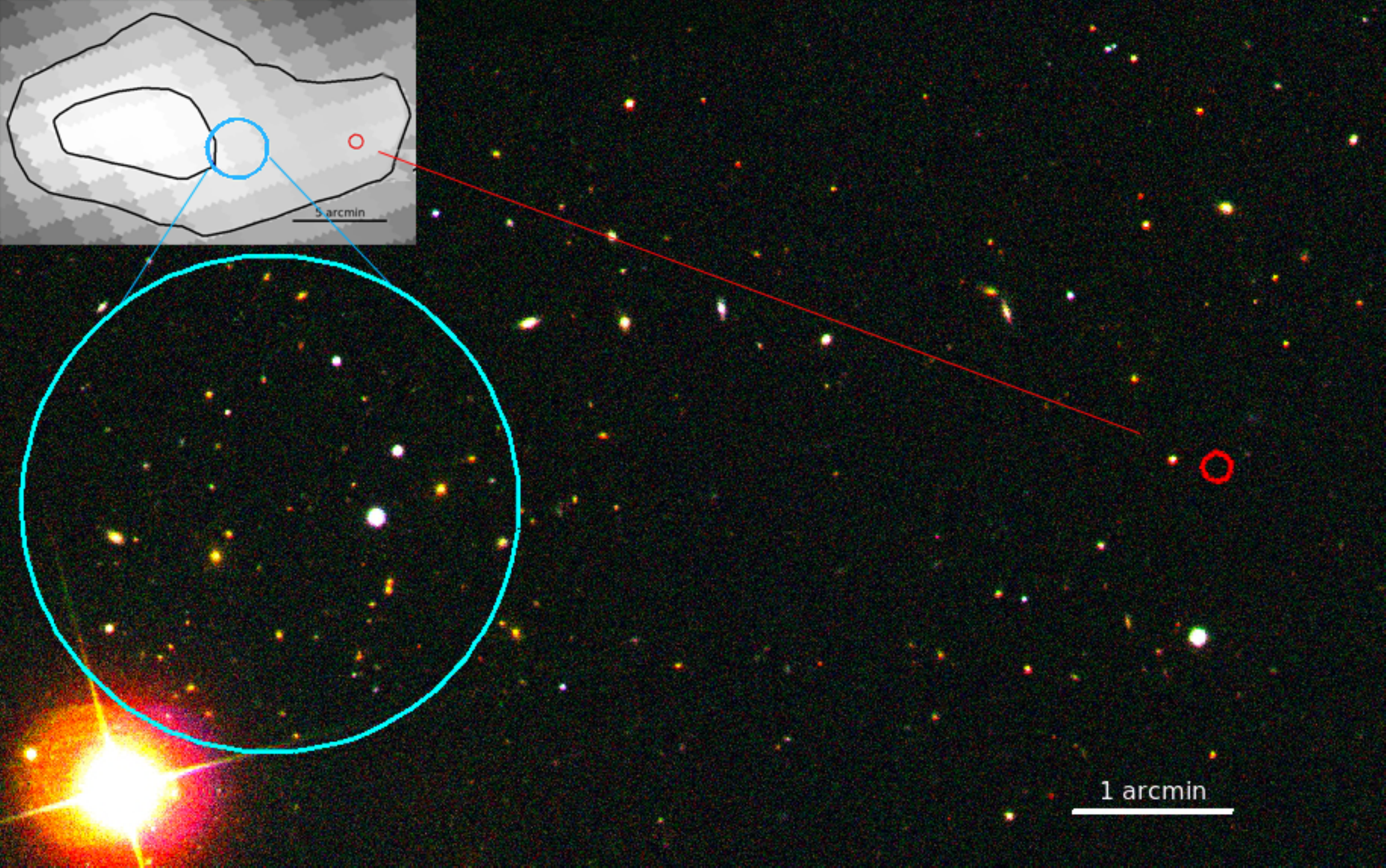}
\caption{The optical counterpart of PSZ2 G231.41+77.48 is located at  $\sim$ 5.7$\arcmin$ from the nominal \Planck\ position (the red small 
circle). The clear shift in the distribution of the SZ signal (seeing in the MILCA contours) supports this association.
}
\label{fig:psz2_1049}
\end{figure}

In addition, in order to obtain a robust confirmation of optical counterpart, we take into account the distance between optical 
centre and the nominal \Planck\ SZ coordinates. The position error predicted for SZ detections in the \Planck\ SZ maps is about 
$2\arcmin$ for targets in the PSZ1 sample \citep{planck2013-p05a, barrena18}. Therefore, optical counterparts are not expected to be found beyond 2.5 times the beam FWHM, which means that the cluster associated with the SZ effect should be closer than $\sim 5\arcmin$ 
from the SZ PSZ1 source coordinates. However, when the clusters are nearby systems ($z<0.2$) their apparent radius may fill a large 
region, and in these cases their centre offsets relative to their SZ position may be larger. Fig.~\ref{fig:sdss_z3} (right panel) shows the
distance of the 51 cluster optical centres to the nominal PSZ2 coordinates. No optical counterparts, except PSZ2 G231.41+77.48 and 
PSZ2 G084.69-58.60,  are found beyond $5\arcmin$ from \Planck\ nominal pointing at $z>0.2$. We include these clusters as 
true counterparts due to its specific shape seen in $y$-map (see Fig.~\ref{fig:psz2_1049} and discussion below).
Fig.~\ref{fig:sdss_z2} shows the final offset distribution of cluster optical centre relative to their \Planck\ SZ position. We find that 68\% of the 37 confirmed cluster sample are enclosed within $3\arcmin.1$, which is consistent with what is expected for SZ \Planck\ target selection \citep{planck2013-p05a, barrena18}.

In Table \ref{tab:newpsz2} we summarize our results for the first subgroup of 37 sources (distance $< 5\arcmin$ or/and inside 1 Mpc and 
richness $>$5), providing the official ID number, taken from the PSZ2 catalogue, the \Planck\ Name, optical 
counterpart coordinates (assumed to be the BCG position or, in the absence of BCG, as approximate geometrical centre of the  
likely members), redshift (photometric and, if available, spectroscopic) and number of spectroscopically confirmed galaxy members. We also 
provide spectroscopic information for 10 sources: 8 have spectroscopic coverage from the SDSS DR12 survey and, for two cases we 
quote the spectroscopic redshift from our long-slit observations using ACAM spectrograph installed at the William Herschel Telescope 
\citep[for a full description of the used instruments set-up see][]{planck2015-XXXVI}. 
Multiple optical counterparts are identified for three SZ sources. We also provide richness information for each cluster. Most of our optical counterparts are clusters included in the catalogue 
of \citet{wen12}. However, we update the position (attending to the BCG identifications) and redshift information. 
All clusters have signal-to-noise ratio (SNR) of the SZ signal (using the MMF1, MMF3 or PwS detection pipelines) between 4.5 and 5.8 
except PSZ2 G237.68+57.83 (SNR=7.4) and PSZ2 G153.56+36.82 (SNR=15.9).

The second subgroup of 17 clusters does not fit the conditions of minimal distance and richness imposed to be considered actual optical
counterparts linked to the SZ emission. This means that this group of clusters are beyond $5\arcmin$ from the nominal \Planck\ SZ centre 
or are not rich enough systems. However, due to the fact that some of the \Planck\ clusters  from the PSZ1 catalogue were already detected at larger distances by the previous teams \citep[see][]{planck2015-XXXVI, barrena18} and that these objects correspond to actual clusters in SDSS DR12 images, we decided to keep these sources for further detailed investigation. So, we quote them as "possible counterparts" and present them in Table \ref{tab:newpsz2_add}.

In the rest of the cases, the cluster identification performed on SDSS DR12 imaging data was not conclusive. Those SZ targets may be associated to areas with strong dust contamination (so producing false SZ detections in the \Planck\ maps), or the corresponding cluster counterparts 
are high redshift clusters, so not visible in the SDSS images due to their low depth. We will re-observe these fields during our follow-up  in order to obtain deep images and confirm either the distant cluster or false SZ detection.

In the following, we describe, as example, results for a few clusters at different redshifts.

\begin{figure}[ht!]
\centering
\includegraphics[width=8.5cm]{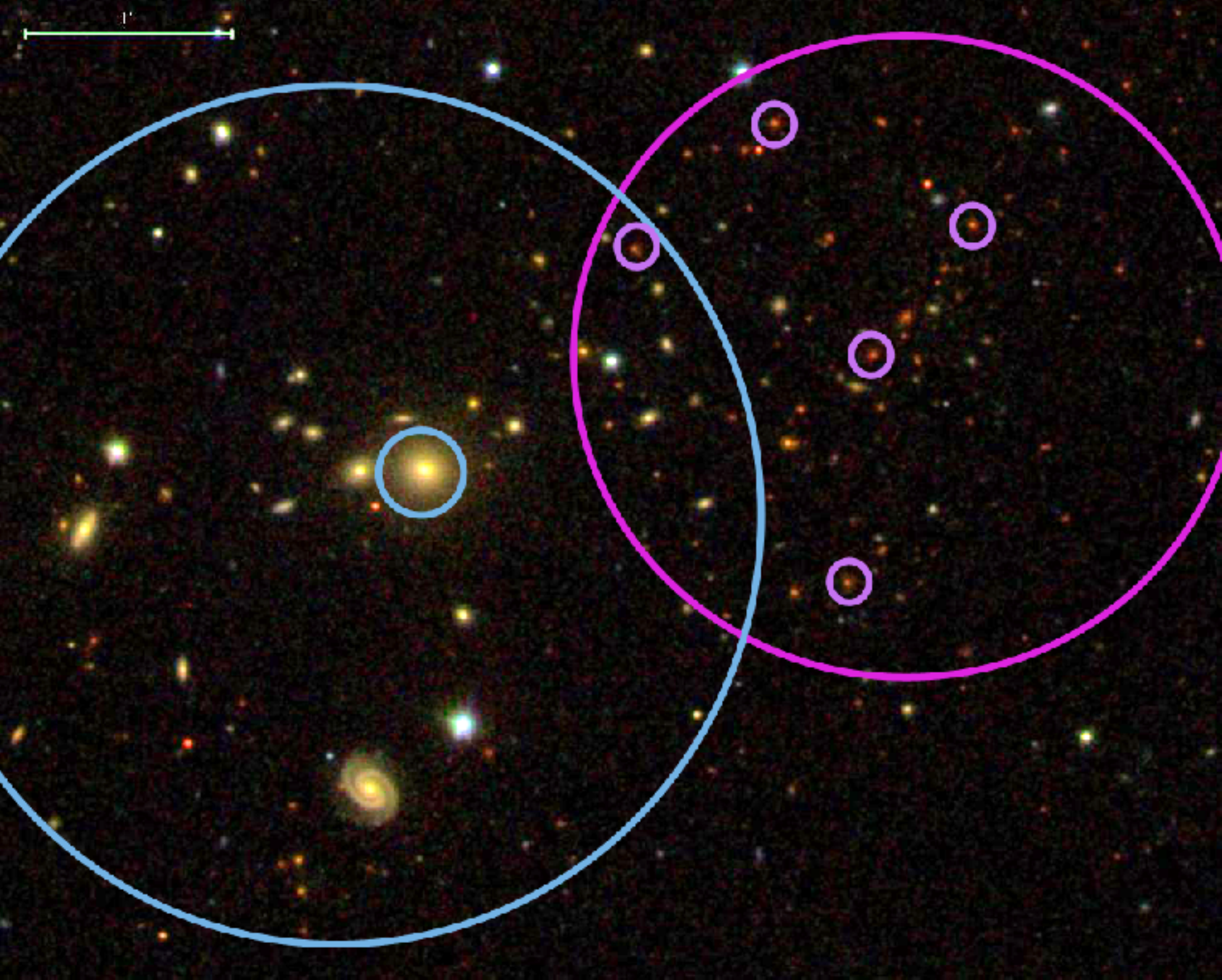}
\caption{RGB image of the SZ source PSZ1 G213.95+68.28. This target shows the presence of 2 possible optical counterparts at different redshifts.
One at low redshift, $z_{\rm spec}$= 0.10 (blue circle), and a second cluster in the background at $z_{\rm spec}$= 0.6 (magenta circle). 
Spectroscopic cluster members are marked with small circles. For both clusters we identify more than 20 spectroscopically confirmed members.
}
\label{fig:psz1_705}
\end{figure}

\paragraph{PSZ2 G069.35$-$15.58} 

This is a low redshift system at $z_{\rm phot} \sim$0.09 (see Fig.~\ref{fig:psz2_293}). We obtain R = 25 for this cluster, which is one of the richness of the clusters studied in this work. We have acquired spectroscopic redshifts for two cluster members using ACAM/WHT long-slit spectroscopy under an International Time  Program  (ITP13)  of  the  ORM, obtaining a mean $z_{\rm spec}$ of 0.095. 

\paragraph{PSZ2 G086.28$+$74.76} 

The importance of visual inspection of RGB images is clearly shown with this example (see Fig.~\ref{fig:psz2_381}, left panel). 
This is a distant cluster at $z_{\rm spec}=$ 0.6995. Due to this high redshift, the cluster galaxy population is very faint. The BCG has 
an apparent magnitude of $r$'=21.63, so the SDSS catalogues only include photometry for a few cluster members, thus the red sequence is completely 
missing in the corresponding colour-magnitude diagrams. For this reason, the richness cannot be estimated. However, by "eye inspection" we are able to 
identify 12 likely cluster members with colours in agreement with a $z_{\rm phot}\sim$0.75$^{+0.07}_{-0.08}$. Moreover, the brightest galaxy 
(probably the BCG of this cluster) has a $z_{\rm spec}=$ 0.6995 in the SDSS DR12 spectroscopic catalogue.

\paragraph{PSZ2 G231.41+77.48} 

Even though this cluster is slightly beyond the limit of our accepted 5$\arcmin$ radius search (5$\arcmin$.7), the MILCA contours confirm 
that this cluster is our optical counterpart to the SZ signal (Fig.~\ref{fig:psz2_1049}).  We found 2 spectra (including the BCG) 
at $z_{\rm spec} \sim$ 0.346. A similar situation appears with PSZ2 G084.69-58.6, which shows a similar shift between the peak of the SZ signal ($y$-map) and the nominal \Planck\ position.

\paragraph{PSZ2 G270.78$+$36.83 and PSZ2 G270.88$+$37.23} 

These two PSZ2 sources are separated by a distance of $\sim$ 25$\arcmin$ (see Fig.~\ref{fig:psz2_381}, right panel). The optical counterpart of 
PSZ2 G270.88+37.23 is a rich cluster (R=15) at $z_{\rm phot} \sim$ 0.23 catalogued by \citet{wen12} as WHL J110504.6-185639. However, 
PSZ2 G270.78+36.83 seems to be a multiple detection, potentially associated with two clusters at different redshifts along  the line of sight and 
almost overlapping in the sky plane. One of these two clusters is a system at $z\sim$ 0.22,
while the second clump is a system at $z_{phot}\sim$ 0.52. Probably both clusters are contributing to the SZ emission, although their separate 
contribution requires further investigation. In total, we find three cases like this, where multiple counterparts are associated with a single 
SZ source. We mark these sources with a special symbol in the Table \ref{tab:newpsz2}.



\section{Results for the \Planck\ PSZ1 catalogue}
\label{sec:psz1}

We also inspect SDSS DR12 data in order to update the spectroscopic redshift information of PSZ1 clusters that were already confirmed photometricaly 
\citep{planck15}. We find 71 cluster fields in the SDSS footprint, but only 48 of them have also spectroscopic coverage. 
We have been able to retrieve spectroscopic information for 33 clusters, namely cases where likely members were
spectroscopically observed. Almost half of the clusters 
have only one or two galaxy members with spectra. However, in five cases we find more than 10 members. 
In all 33 identifications, spectroscopic redshifts were in good agreement with the previous photometric data. No significant 
discrepancy was detected. The main reason for this is that an important fraction of these previous identifications belong to 
clusters retrieved through the cross-correlation of the original PSZ1 catalogue with the SDSS-III photometric cluster catalogue 
by \citet{wen12}.

Additionally, we inspect a few still unconfirmed PSZ1 clusters inside the SDSS DR12 data and we have been able to identify the optical 
counterparts for the \Planck\ SZ source PSZ1 G213.95+68.28. There are two possible identifications for this target, both of which are
spectroscopically confirmed (see Fig.~\ref{fig:psz1_705}). This is a new case of multiple detection with two clear clusters at 
different redshifts along the line of sight. The low-z cluster is at $z\sim$0.10, showing about 50 likely members, 10 of them being
spectroscopically confirmed. This low-z cluster coincides with a compact cluster in the background, at $z\sim$0.601. 
In fact, PSZ1 G213.95+68.28 is a rich structure with five cluster members at 0.595$<z_{\rm spec}<$0.603 and 12 likely members 
showing a $z_{\rm phot}\sim$0.6$^{+0.1}_{-0.1}$. However, both systems are inside the peak of the \Planck\ SZ detection in the 
$y$-maps and thus probably both are contributing to the SZ emission.

The identification of the BCG in clusters at low redhisft is straightforward. However, in clusters at $z>0.5$ the BGC is not so clear. 
In general, clusters at high redshift are young and not so evolved structures, and have less luminous and smaller BCGs than clusters at lower $z$ \citep{ascaso11}. 
In such cases, several galaxies are typically found with very similar magnitudes in the core of the clusters. Therefore, for these high redshift clusters we provide 
the coordinates of the average position of all spectroscopically confirmed members. 

Our new spectroscopic redshifts for the PSZ1 sources are presented in Table \ref{tab:inpsz1}.


\section{Conclusions}
\label{sec:conclusions}


We carried out a detailed study of \Planck\ PSZ2 targets using optical information from SDSS DR12.
The technique presented here also combines the information from the Compton $y$-maps. We implement restrictions concerning the optical richness 
and distance from the nominal \Planck\ pointing in order to validate the cluster counterparts. Together, this procedure allows for a more robust 
identification of optical counterparts compared to simply cross-matching with existing SDSS cluster catalogues that have been constructed from earlier SDSS Data Releases.

Using this method we confirm 37 new \Planck\ PSZ2 clusters. We also identified 17 clusters more in the SDSS images potentially 
associated with the corresponding SZ emission. However, this cluster sample does not follow completely the distance/richness requirements. 
These clusters could contribute somehow to the corresponding SZ emission but further investigations (mainly multiobject spectroscopy to obtain dynamical masses) are needed. 

For 11 of the 37 new clusters we present spectroscopic redshifts (for 9 clusters from SDSS DR12 database, and 2 clusters from our 
long-slit observations using ACAM/WHT). Three SZ sources present multiple optical counterparts. For these cases, we find several clusters 
along the line of sight of the SZ signal.


Finally, we present updated spectroscopic information for 34 \Planck\ PSZ1 sources, 33 of them being previously confirmed photometrically. 
Additionally, we validate the cluster PSZ1 G213.95+68.28. This is a PSZ1 source showing two clusters along the line-of-sight at $z=0.101$ and $z=0.601$.

The work presented here is part of a more general programme to optically validate \Planck\ SZ sources trying to determine the 
purity and efficiency of cluster detections in \Planck\ SZ maps.


\begin{acknowledgements}
We thank Monique Arnaud for helpful comments.
This research has made use of the following databases:  the
SZ-Cluster Database operated by the Integrated Data and Operation Center (IDOC)
at the IAS under contract with CNES and CNRS; and the SDSS.
Funding for the Sloan Digital Sky Survey (SDSS) has been provided by the Alfred
P. Sloan Foundation, the Participating Institutions, the National Aeronautics
and Space Administration, the National Science Foundation, the U.S. Department
of Energy, the Japanese Monbukagakusho, and the Max Planck Society. 
This research has been carried out with telescope time awarded by the CCI International
Time Programme at the Canary Islands observatories (program ITP13-8).
The article is based on observations carried out at the William Herschel Telescope
operated on the island of La Palma by the ISAAC Newton Group of Telescopes in the Spanish ORM of the IAC.
This work has been partially funded by the Spanish Ministry of Economy and
Competitiveness (MINECO) under the projects ESP2013-48362-C2-1-P and AYA2014-60438-P.
AS and RB acknowledge financial support from the Spanish Ministry
of Economy and Competitiveness (MINECO) under the 2011 Severo Ochoa Program MINECO SEV-2011-0187.
HL was supported by Estonian Research Council (ETAg) projects PUT1627,
IUT26-2 and IUT40-2, and by the Centre of Excellence "Dark side of the
Universe" (TK133) financed by the European Union through the European
Regional Development Fund.
RFJvdB acknowledges support from the European Research 
Council under FP7 grant number 340519.

\end{acknowledgements}

\bibliographystyle{aa}

\clearpage
\begin{landscape}
\begin{table} 
\begin{center}
\caption[]{\label{tab:newpsz2}List of new 37 PSZ2 clusters confirmed using the SDSS DR12 data. The first two columns are the index number and the official name of the SZ target in the PSZ2 catalogue. The J2000 coordinates (column 3 and 4) correspond to the BCG position or geometrical centre (labeled as Cen=1 or Cen=2 in column 5, respectively) of the identified 
counterpart. Column 6 shows the photometric redshift of the cluster. Columns 7, 8 and 9 show (if available) the mean spectroscopic redshift, 
spectroscopic redshift of the BCG (in the case of absence of the apparent BCG we write "-1") and the number of galaxies with spectroscopic 
measurements. Column 10 provides the r magnitude (corrected for extintion) of the BCG. Columns 11 presents optical richness of the cluster and its error, where
possible (see Sec. 4.4 for details). Last column includes the source name in \citet{wen12} }
\begin{tabular}[h]{cccccccccccc}
    \hline \hline
ID & Planck Name & R. A. & Decl. & Cen. & $z_{\rm phot}$ & $z_{\rm spec}$ & $z_{\rm spec}(BCG)$ & N$_{\rm spec}$ & $r_{\rm mag}(BCG)$ & R &comments \\
 \hline \hline
34  & PSZ2 G009.04+31.09 & 16:18:26.70 & -04:11:11.06 & 1 &   0.25 & 0       & 0 & 0 & 17.42 & 31 $\pm$ 5.5 & WHL J161826.7-041111  \\
68  & PSZ2 G021.02-29.04 & 20:20:28.21 & -22:25:14.92 & 1 &   0.32 & 0.305$^{a}$ & 0 & 2$^{a}$ & 17.74 & 15 $\pm$ 3.8 & WHL J202028.2-222515\\
115 & PSZ2 G032.31+66.07 & 14:37:24.97 & +24:23:54.96 & 2 &   0.62 & 0.609   & -1 & 6 & - & & WHL J143725.0+242355 \\
161 & PSZ2 G044.21+52.13 & 15:42:50.44 & +27:49:52.90 & 1 &   0.39 & 0.378   & 0.3775 & 2 & 18.91 & 12 $\pm$ 3.4 & WHL J154250.4+274953 \\
176 & PSZ2 G045.96-26.94 & 20:50:01.00 & -01:35:24.30 & 1 &   0.20 & 0 & 0   & 0 & 17.25 & 10 $\pm$ 3.1&   \\
277 & PSZ2 G066.34+26.14 & 18:01:06.52 & +39:52:06.73 & 1 &   0.63 & 0 & 0   & 0 & 19.96 & 10 $\pm$ 3.1&  WHL J180111.8+395138  \\
279 & PSZ2 G066.59-58.51 & 23:07:11.30 & -07:31:43.15 & 1 &   0.36 & 0 & 0   & 0 & 18.88 & 7 $\pm$ 2.6  & WHL J230711.3-073143  \\
284 & PSZ2 G067.21-20.49 & 21:13:28.88 & +18:03:27.47 & 1 &   0.31 & 0 & 0   & 0 & 17.45 & 5 $\pm$ 2.2 & WHL J211328.9+180327 \\
293 & PSZ2 G069.35-15.58 & 21:02:41.68 & +22:43:13.63 & 1 &   0.09 & 0.094$^{a}$ & 0.0958$^{a}$ & 2$^{a}$ & 14.87 & 25 $\pm$ 5 & \\ 
294 & PSZ2 G069.39+68.05 & 14:21:38.35 & +38:21:17.78 & 2 &   0.71 & 0.762  & 0.762 & 1 & 21.23 &   & ClG-J142138.3+382118$^{b}$\\
295 & PSZ2 G069.47-29.06 & 21:46:02.16 & +14:01:25.90 & 1 &   0.36 & 0      & 0 & 0 & 19.49 & 11 $\pm$ 3.3& \\
330 & PSZ2 G077.67+30.59 & 17:46:50.85 & +50:31:11.88 & 1 &   0.23 & 0      & 0 & 0 & 18.07 & 6 $\pm$ 2.4 & WHL J174650.9+503112 \\ 
371 & PSZ2 G084.69-58.60 & 23:36:37.52 & -01:27:52.34 & 1 &   0.19 & 0.185  & 0 & 2 & 16.25 & 7 $\pm$ 2.6 & WHL J233637.5-012752  \\ 
381 & PSZ2 G086.28+74.76 & 13:37:53.87 & +38:54:06.82 & 1 &   0.8  & 0.699  & 0.6995 & 1 & 21.63 &  & \\ 
394 & PSZ2 G087.39-34.58 & 22:49:09.53 & +19:44:30.50 & 2 &   0.7  & 0.76   & -1 & 1 & - &  & \\
424 & PSZ2 G093.41-16.26 & 22:24:07.25 & +37:58:30.46 & 1 &   0.26 & 0      & 0  & 0 & 17.37 & 7 $\pm$ 2.6 & WHL J222407.2+375831 \\
432 & PSZ2 G094.31-11.31 & 22:12:58.51 & +42:35:02.26 & 1 &   0.24 & 0      & 0  & 0 & 16.62 & 13 $\pm$ 3.6 & \\
483 & PSZ2 G100.22+33.81 & 17:13:45.53 & +69:21:48.07 & 2 &   0.62 & 0      & -1 & 0 & - &  & \\
620 & PSZ2 G125.84-18.72 & 01:06:55.84 & +44:06:05.81 & 2 &   0.19 & 0      & -1 & 0 & - & 16 $\pm$ 4 & WHL J010709.2+440918.  \\
624 & PSZ2 G126.36-19.11 & 01:09:19.54 & +43:37:40.78 & 1 &   0.22 & 0      & 0 & 0 & 16.37 & 10 $\pm$ 3.1 & WHL J010919.5+433741 \\
628 & PSZ2 G126.72-21.03 & 01:10:25.02 & +41:41:18.81 & 1 &   0.22 & 0      & 0 & 0 & 15.83 & 5 $\pm$ 2.2& WHL J011025.0+414119 \\
644$^{c}$ & PSZ2 G130.64+37.16 & 10:47:45.54 & +77:59:56.67 & 1 & 0.44 &  0 & 0 & 0 & 19.98 & 6 $\pm$ 2.4& WHL J104745.5+775957 \\
644$^{c}$ & PSZ2 G130.64+37.16 & 10:46:29.31 & +78:07:44.06 & 1 & 0.24 &  0 & 0 & 0 & 17.10 & 7 $\pm$ 2.6 & WHL J104629.2+780744 \\
646 & PSZ2 G131.15-14.72 & 01:38:42.22 & +47:22:35.27 & 2 &   0.24 & 0      & -1 & 0 & - & 31  $\pm$ 5.5 & WHL J013846.1+472236. \\ 
681 & PSZ2 G139.00+50.92 & 11:20:16.97 & +63:14:59.87 & 2 &   0.6  & 0      & -1 & 0 & - &  & \\
732 & PSZ2 G150.64-14.21 & 03:17:04.20 & +40:41:33.22 & 1 &   0.22 & 0      & 0 & 0 & 16.86 & 10 $\pm$ 3.1& WHL J031704.2+404133  \\
810 & PSZ2 G171.08-80.38 & 01:21:53.35 & -20:33:26.69 & 1 &   0.33 & 0      & 0 & 0 & 17.36 & 6 $\pm$ 2.4 & WHL J012153.4-203327 \\ 
831 & PSZ2 G177.03+32.64 & 08:13:20.93 & +43:12:36.83 & 1 &   0.16 & 0.1428 & 0.143 & 6 & 16.85 & 20 $\pm$ 4.4& WHL J081320.9+431237 \\ 
836 & PSZ2 G179.45-43.92 & 03:19:18.34 & +02:05:35.43 & 1 &   0.34 & 0 & 0  & 0 & 16.91 & 5 $\pm$ 2.2& WHL J031918.3+020535  \\ 
863 & PSZ2 G186.50-13.45 & 05:11:52.85 & +16:13:36.29 & 1 &   0.25 & 0 & 0  & 0 & 16.6 & 9 $\pm$ 3& WHL J051152.9+161336  \\ 
1049 & PSZ2 G231.41+77.48 & 12:00:26.54 & +22:34:19.55 & 1 &  0.35 & 0.346  & 0.3469 & 2  & 18.55 & 10 $\pm$ 3.1 & WHL J120026.5+223420 \\  
1074 & PSZ2 G237.68+57.83 & 10:53:17.80 & +10:52:37.13 & 1 &  0.7  & 0 & -1 & 0 & - &  & \\
1244 & PSZ2 G269.02+46.30 & 11:19:07.43 & -10:22:57.81 & 2 &  0.12 & 0 & -1 & 0 & - & 8 $\pm$ 2.8 &\\ 
1254$^{c}$ & PSZ2 G270.78+36.83 & 11:04:06.99 & -19:13:47.88 & 1 & 0.22 & 0 & 0 & 0 & 16.88 & 12 $\pm$ 3.4 & WHL J110407.0-191348 \\ 
1254$^{c}$ & PSZ2 G270.78+36.83 & 11:04:21.03 & -19:14:18.12 & 1 & 0.52 & 0 & 0 & 0 & 19.59 & 18 $\pm$ 4.2 & WHL 110417.9-191409 \\ 
1255 & PSZ2 G270.88+37.23 & 11:05:04.58 & -18:56:38.66 & 1 & 0.23 & 0 & 0   & 0 & 16.62 & 15 $\pm$ 3.8 & WHL J110504.6-185639 \\ 
1262$^{c}$ & PSZ2 G271.53+36.41 & 11:05:25.72 & -19:54:19.84 & 1 & 0.48 & 0 & 0 & 0 & 17.84 & 5 $\pm$ 2.2 & WHL J110525.7-195420 \\ 
1262$^{c}$ & PSZ2 G271.53+36.41 & 11:05:19.71 & -19:59:15.61 & 2 & 0.50 & 0 & -1 & 0 & - & 6 $\pm$ 2.4& WHL J110519.6-195852\\ 
1465 & PSZ2 G310.81+83.91 & 12:55:18.02 & +21:02:31.22 & 1 &  0.44 & 0.429  & 0.4288 & 2 & 19.54 & 5 $\pm$ 2.4& PSZ1 G311.76+83.91\\
1548 & PSZ2 G328.96+71.97 & 13:23:27.81 & +10:46:43.71 & 2 &  0.09 & 0.092  & -1 & 10 & - & 10 $\pm$ 3.1&  \\
\hline   
  \end{tabular}
\end{center}
\small 
$^{a}$ Spectroscopic redshift obtained from optical follow-up observations with the WHT telescope. No spectroscopic SDSS data available.\\
$^{b}$ additional analysis presented in \citet{bud15} \\
$^{c}$ Multiple optical detection. \\
\end{table}
\clearpage
\end{landscape}


\begin{table*} 
\caption[]{\label{tab:newpsz2_add} List of 17 optical clusters 
"potentially associated" with the corresponding PSZ2 target. These targets do not fit the selection requirements of distance/richness to the \Planck\ SZ detection. We quote them as "possible counterparts" until further investigations are performed. Columns follow the same scheme of Table \ref{tab:newpsz2}.
}
\begin{tabular}[h]{ccccccccccc}
    \hline \hline
ID & Planck Name & R. A. & Decl. & Cen. & $z_{\rm phot}$ & $z_{\rm spec}$ & $z_{\rm spec}BCG$ & N$_{\rm spec}$ & $r_{\rm mag}$ &  comments \\
 \hline \hline
126 & PSZ2 G036.36+16.01 & 17:59:42.69 & +10:07:43.70 & 1 &   0.13 & 0.128$^{a}$ & 0.1278$^{a}$ & 3$^{a}$ & 16.21 &  \\
327 & PSZ2 G076.55+60.29 & 14:52:00.51 & +44:31:21.31 & 2 &   0.29 & 0.287 & -1 & 4 & 0 & WHL J145206.4+443235 \\  
389 & PSZ2 G086.85-81.39 & 00:30:15.21 & -20:06:24.93 & 1 &   0.25 & 0     & 0 & 0 & 17.28 & WHL J003015.2-200625  \\
421 & PSZ2 G092.69+59.92 & 14:26:08.53 & +51:14:14.15 & 2 &   0.50 & 0.461 & -1 & 3 & 0 &   \\
465 & PSZ2 G098.62+51.76 & 14:50:05.70 & +59:19:51.91 & 1 &   0.31 & 0.298 & 0.2985 & 1 & 18.16 & WHL J145005.7+591952  \\ 
546 & PSZ2 G112.69+33.37 & 16:19:49.38 & +79:06:23.53 & 1 &   0.51 & 0     & 0 & 0 & 19.12 & WHL J161949.3+790624  \\ 
667 & PSZ2 G136.02-47.15 & 01:28:23.61 & +14:41:13.60 & 1 &   0.50 & 0.465 & 0.4649 & 1 & 19.00 & WHL J012823.6+144114  \\ 
673 & PSZ2 G137.24+53.93 & 11:40:59.55 & +61:07:07.04 & 1 &   0.48 & 0.473 & 0.4769 & 2 & 18.98 & WHL J114059.5+610707 \\ 
690 & PSZ2 G141.98+69.31 & 12:12:40.63 & +46:21:23.07 & 1 &   0.70 & 0.714 & 0.7132 & 2 & 21.10 & WHL J121240.6+462123\\ 
714 & PSZ2 G146.16-48.92 & 01:52:40.99 & +11:14:30.90 & 2 &   0.49 & 0.491 & -1 & 9 & 0 & \\ 
739 & PSZ2 G152.40+75.00 & 12:13:19.22 & +39:46:26.91 & 1 &   0.42 & 0.453 & 0.4513 & 6 & 19.16 & WHL J121319.2+394627 \\ 
744 & PSZ2 G153.56+36.82 & 08:44:21.57 & +62:17:48.40 & 2 &   0.12 & 0.122 & -1 & 10 & - &  \\
812 & PSZ2 G171.48+16.17 & 06:37:43.56 & +43:48:59.12 & 1 &   0.28 & 0     & 0 & 0 & 17.15 & WHL J063743.6+434859 \\ 
916 & PSZ2 G202.61-26.26 & 04:59:50.17 & -03:16:47.52 & 1 &   0.23 & 0 & 0 & 0 & 17.12 & WHL J045950.2-031647  \\ 
917 & PSZ2 G202.66+66.98 & 11:07:30.89 & +28:51:01.10  & 1 &  0.48 & 0.483 & 0.4814 & 3 & 19.07 & WHL J110730.9+285101 \\ 
920 & PSZ2 G203.32+08.91 & 07:05:56.53 & +12:30:33.66 & 1 &   0.27 & 0 & -1 & 0 & 0 & WHL J070556.5+123034 \\ 
1510 & PSZ2 G320.94+83.69 & 13:00:05.73 & +21:01:28.26 & 1 &  0.45 & 0.462 & 0.4612 & 4 & 19.12 & WHL J130005.7+210128 \\ 
\hline   
  \end{tabular}
\small 
$^{a}$ Spectroscopic redshift obtained from optical follow-up observations with the WHT telescope. No spectroscopic SDSS data available. \\
\end{table*}

 \begin{table*}
\caption[]{\label{tab:inpsz1} List of confirmed clusters from PSZ1 with updated spectroscopic redshift. 
The first two columns provide the index number and official name of the cluster in the PSZ1 catalogue. 
The J2000 coordinates correspond to the BCG or geometrical centre of the identified counterpart. 
Columns 5 and 6 show the new cluster redshifts and the number of galaxy members with spectroscopic measurements.} 
\begin{center}
 \begin{tabular}[h]{cccccc}
    \hline
    \hline
ID& Planck Name & R. A.& Decl. & $z_{\rm spec}$ & N$_{\rm spec}$ \\
 \hline 
\hline
52&  PSZ1 G020.82$+$38.03&   16:15.05.44& $+$07:34:55.38&    0.338&   4\\
97&  PSZ1 G035.55$+$34.15&   16:50:35.73& $+$16:49:31.27&    0.348&   3 \\
142& PSZ1 G046.98$+$66.62&   14:37:40.29& $+$30:12:00.27&    0.339&   3 \\  
143& PSZ1 G047.44$+$37.39&   16:50:20.40& $+$26:58:21.42&    0.232&   1 \\
257& PSZ1 G078.39$+$46.13&   16:09:01.48& $+$50:05:11.85&    0.400&   1 \\
260& PSZ1 G079.65$+$45.60&   16:10:54.67& $+$51:13:35.91&    0.392&   2 \\
299 & PSZ1 G087.32$+$50.92&  15:26:13.83& $+$54:09:16.6 &  0.473 &   4\\
307& PSZ1 G090.63$+$33.50&   17:27:46.53& $+$61:30:27.82&    0.312&   1 \\
312$^{a}$& PSZ1 G091.81$-$26.97&   22:45:27.74& $+$28:09:00.78&    0.344&   15 \\
321& PSZ1 G093.15$+$43.20&   16:05:15.53& $+$61:04:42.54&    0.384&   3 \\
357$^{b}$& PSZ1 G099.84$+$58.45&   14:14:47.18&  $+$54:47:03.57&   0.620 & 2\\
392& PSZ1 G107.67$-$39.80&   00:01:17.36&  $+$21:31:22.04&  0.411&  18\\
404& PSZ1 G110.08$-$70.23&   00:33:53.14&  $-$07:52:10.35&  0.302&  3\\
410 &PSZ1 G112.38$-$32.88&   00:10:51.39& $+$29:09:40.15&  0.332&  19\\
418& PSZ1 G114.34$-$60.16&   00:34:28.16& $+$02:25:22.61&  0.382&    3 \\
427&PSZ1 G116.48$-$44.47&    00:32:08.22& $+$18:06:25.28&  0.374&  11\\
441& PSZ1 G120.14$-$44.43&   00:42:36.61& $+$18:25:33.25&  0.281 & 3\\
448 &PSZ1 G122.98$-$35.52&   00:51:38.60& $+$27:19:59.89&  0.358&   6\\
461& PSZ1 G126.00$-$49.62&   00:59:36.76& $+$13:10:20.84&  0.501&  8\\
487& PSZ1 G135.24$+$65.43&   12:19:12.20& $+$50:54:35.04&  0.552& 7\\
505& PSZ1 G140.46$+$54.27&   11:30:08.72& $+$59:46:32.15&    0.169&    6\\
576& PSZ1 G165.94$+$50.48&   10:01:38.77& $+$50:00:53.59&    0.172&    2 \\
579& PSZ1 G166.61$+$42.12&   09:09:38.69& $+$51:36:15.13&    0.215&    3 \\
660& PSZ1 G198.50$+$46.01&   09:30:51.19& $+$28:47:00.26&    0.296&    7 \\
696$^{c}$& PSZ1 G212.51$+$63.18& 10:52:52.73&  $+$24:14:00.14 & 0.529 & 5 \\  
699& PSZ1 G212.80$+$46.65&   09:44:42.87& $+$19:27:59.64&    0.367&    1 \\
701& PSZ1 G213.27$+$78.35&   11:59:31.29& $+$26:14:38.73&    0.306&    3 \\
705$^{d}$& PSZ1 G213.95$+$68.28 &   11:14:41.68& $+$24:34:50.65  &  0.101 & 10\\
705$^{d}$& PSZ1 G213.95$+$68.28 &   11:14:32.15&  $+$24:35:24.06 & 0.601 & 5 \\
719& PSZ1 G217.35$+$58.14&   10:34:39.66& $+$20:32:04.52&    0.476&    2 \\
743& PSZ1 G223.80$+$58.50&   10:41:09.64& $+$17:30:35.08&    0.337&    2 \\
760& PSZ1 G226.65$+$28.43&   08:56:20.59& $+$01:46:49.38&    0.724&    1 \\
795& PSZ1 G236.86$+$66.33&   11:21:55.15& $+$15:48:05.16&    0.675&    1 \\
1093& PSZ1 G311.76$+$83.91&   12:55:18.02& $+$21:02:31.22&    0.429&    2 \\
1159& PSZ1 G332.30$+$72.17&   13:26:43.58& $+$11:17:04.22&    0.089&    14 \\
\hline     
  \end{tabular}\\
\end{center}
\small 
$^{a}$ centre of the main group. This system consists of two groups at $z$ 0.344 and 0.339 \\
$^{b}$ centre of the main rich cluster. The second nearby cluster has $z_{\rm spec}$ of 0.579 \\
$^{c}$ updated coordinates (centre of concentration)\\
$^{d}$ double detection\\
\end{table*}




\end{document}